\documentclass[a4paper,11pt]{article} 

\usepackage{graphicx}          % Include this line if your 
\usepackage[toc,page]{appendix}
\usepackage{color}
\usepackage{dsfont}
\usepackage{extarrows}
\usepackage{enumerate}
\usepackage{amsmath}    % need for subequations
\usepackage{amssymb}
\usepackage{bbm}
\usepackage{graphicx}   % need for figures
\usepackage{mathrsfs}
\usepackage{verbatim}   % useful for program listings
\usepackage{subfigure}  % use for side-by-side figures
\usepackage{hyperref}   % use for hypertext links, including those to external documents and URLs
\newtheorem{theorem}{Theorem}[section]
\newtheorem{corollary}[theorem]{Corollary}
\newtheorem{lemma}[theorem]{Lemma}

\newtheorem{definition}[theorem]{Definition}
\newtheorem{remark}[theorem]{Remark}

\newcommand{\Tr}{{\rm Tr}}

\title{\Large \bf
Dissipative Feedback Switching for Quantum Stabilization
}
\date{\vspace{-5ex}}

\author{Weichao Liang,
\thanks{
{\small Department of Information Engineering, University of Padua, 6B, Via Gradenigo, 35131 Padova, Italy;
(weichao.liang@dei.unipd.it).}}%
\and 
Tommaso Grigoletto,
\thanks{
{\small Department of Information Engineering, University of Padua, 6B, Via Gradenigo, 35131 Padova, Italy; (tommaso.grigoletto@phd.unipd.it).}}
\and 
Francesco Ticozzi,
\thanks{
{\small Department of Information Engineering, University of Padua, 6B, Via Gradenigo, 35131 Padova, Italy; (ticozzi@dei.unipd.it).}}
}

\begin{document}

\maketitle

\begin{abstract}                          % Abstract of not more than 200 words.
{Switching controlled dynamics allows for fast, flexible control design methods for quantum stabilization of pure states and subspaces, which naturally include both Hamiltonian and dissipative control actions. A novel approach to measurement-based, dissipative feedback design is introduced, and extends the applicability of switching techniques with respect to previously proposed ones, as it does not need stringent invariance assumptions, while it still avoids undesired chattering or Zeno effects by modulating the control intensity. When the switching dynamics do leave the target invariant, on the other hand, we show that exponential convergence to the target can be enforced without modulation, and switching times that can be either fixed or stochastic with hysteresis to avoid chattering. The effectiveness of the proposed methods is illustrated via numerical simulations of simple yet paradigmatic examples, demonstrating how switching strategies converge faster than open-loop engineered dissipation.}
\end{abstract}

\section{Introduction}
\label{sec:introduction}

Suitable engineering and control techniques for quantum systems are needed for the development of reliable quantum information processing devices \cite{altafini2012modeling}. Feedback control methods, which dominate classical applications, present unique challenges in the quantum domain, as measured systems exhibit stochastic evolutions: a rigorous formalism for their treatment has been developed hinging on stochastic differential equations that describe quantum conditional dynamics, or filtering equations 
\cite{hudson1984quantum,belavkin1989nondemolition,bouten2007introduction,barchielli2009quantum}. A typical task is that of  stabilizing a target state of interest, and different design techniques have been employed to this aim. These include output-feedback methods (also known as Markovian feedback) \cite{ticozzi2008quantum,ticozzi2009analysis}, and filtering-based feedback methods \cite{van2005feedback,mirrahimi2007stabilizing,ticozzi2012stabilization,liang2019exponential,liang2021robustness}.
{Additionally, these control techniques have been extended to scenarios involving time delays~\cite{kashima2009control,szigeti2013robustness}. }
While the dynamics is that of a monitored open system, the control typically enters the dynamics as a perturbation of the system Hamiltonian (a so-called coherent control action). Dissipative resources have been so far mostly used in open-loop control strategy, or as an always-on control action complemented by a purely Hamiltonian feedback \cite{ticozzi2012stabilization}. { However, the possibility of using engineered dissipation for various tasks of interest in quantum information has been explored and offers interesting advantages \cite{dissipativeQC,dissipativeencoding}.

In this spirit, the potential of dissipative switching control has been investigated starting in \cite{scaramuzza2015switching,grigoletto2021stabilization}, considering not only coherent but also dissipative control resources, with the latter obtained as measurements and controlled interaction with suitably engineered fields and quantum environments. It is worth mentioning that most available global stability results already employ switching strategies \cite{ge2012non,wen2022global} to destabilize the undesired equilibria: in our approach we consider switching not just as a technical necessity, but as a true resource.}
In \cite{scaramuzza2015switching}, the switching laws are based on semigroup (also referred to as Lindblad, or master equation) dynamics, that correspond to expected average state of a quantum stochastic filtering equation, and can thus be computed off-line. On the other hand, in \cite{grigoletto2021stabilization} the switching is based on a real-time estimate of the current state of the system, proposing a stabilizing switching feedback law with fixed, sufficiently small dwell times. 
{These results have been obtained under the assumption that the target is invariant for all controlled dynamics, consistently with the typical settings for classical switched systems. Under this assumption, the main role of switching is that of improving the speed of convergence to the target by selecting the locally-optimal generator, as shown in previous works \cite{grigoletto2021stabilization,scaramuzza2015switching}.}

% Introduction to the results of
{
The main contributions of this paper are twofold:\\
(i) Maintaining the invariance assumption, we are able to improve the proof of convergence to show {\em global exponential stability} (GES), ensuring a convergence that is both fast and more robust \cite{liang2019exponential,liang2021robustness}, while providing estimates for the convergence rate. 
We also generalize the approach from deterministic switching with fixed dwell times to stochastic switching times. In doing this, it is crucial to prevent chattering and the so-called Zeno effect - where the switching intervals become infinitesimal in the asymptotic limit. To this aim, we employ a switching law with hysteresis, inspired by \cite{liberzon2003switching,wu2013stability,teel2014stability}.
 
(ii) We propose a novel control design approach, where modulating the intensity of the switched generators allows us to surpass the stringent invariance conditions needed in the existing methods.  By suitably reducing the intensity of the control close to the target, we are able to guarantee almost-sure {\em global asymptotic stability} (GAS) of the target.  This extends significantly the applicability of switching strategies and allows to include many experimental settings where, for example, a feedback use of a destabilizing Hamiltonian is instrumental to stabilization \cite{mirrahimi2007stabilizing,ticozzi2012stabilization}, as well as situations in which dissipative actions can be turned on and off. Without invariance conditions, only a weak result about practical convergence in expectation had been derived in \cite{grigoletto2021stabilization}.}

 In order to further clarify the contributions of the paper and allow for a quick comparison, we report in Table \ref{table} the existing and the new strategies, together with the type of stochastic stability we can guarantee,  GAS or GES.
 
%  \begin{table}
% \tiny
% \caption{\footnotesize \bf Switching strategies and Convergence Type}

% \label{table}
% \begin{tabular}{l||l|l} 
% {\bf Chattering avoidance} & {\bf Average-based } & {\bf Measurement-based}
% \\
%   {\bf and Assumptions} & {\bf switching } & {\bf feedback switching } \\
%     \hline\hline
%     Fixed dwell time & GAS: \cite[Thm 2]{scaramuzza2015switching} & GAS: \cite[Thm 2]{grigoletto2021stabilization}\\
%     {\bf A1}: invariance & GES: Corollary~\ref{Cor:GES_Dwell} & GES: Theorem~\ref{Thm:GES_dwell} \\     
%     \hline
%     Hysteresis switching & GES: Theorem~\ref{Thm:GES_average} & GES: Theorem~\ref{Thm:GES_state} \\
%      {\bf A1}: invariance & & \\ 
%    \hline
%    Modulating Lindbladian & - & GAS: Theorem \ref{Thm:GAS}
%    \\
%      {\bf A2}: no invariance & & \\ 
%    \hline
%    %\multicolumn{3}{p{251pt}}{Stabilization towards target subspaces of open quantum system by state dependent switching.}
% \end{tabular}
% \end{table}

\begin{table}
\tiny
\caption{\footnotesize \bf Switching strategies and Convergence Type}

\label{table}
\begin{tabular}{p{28mm}||p{22mm}|p{22mm}} 
{\bf Chattering avoidance and Assumptions} & {\bf Average-based } {\bf switching } & {\bf Measurement-based switching }\\
    \hline\hline
    Fixed dwell time\newline {\bf A1}: invariance & Law: $\sigma_2$\newline GAS: \cite[Thm 2]{scaramuzza2015switching} GES: Corollary~\ref{Cor:GES_Dwell} & Law: $\sigma_3$ \newline GAS: \cite[Thm 2]{grigoletto2021stabilization} GES: Theorem~\ref{Thm:GES_dwell}\\  
    \hline
    Hysteresis switching \newline {\bf A1}: invariance & Law: $\sigma_1$ \newline GES: Theorem~\ref{Thm:GES_average} & Law: $\sigma_4$\newline GES: Theorem~\ref{Thm:GES_state} \\
   \hline
   Modulating Lindbladian \newline {\bf A2}: no invariance & - & Law: $\sigma_5$ \newline GAS: Theorem \ref{Thm:GAS}\\
   \hline
   %\multicolumn{3}{p{251pt}}{Stabilization towards target subspaces of open quantum system by state dependent switching.}
\end{tabular}
\end{table}

 The paper is organized as follows: Section \ref{sec:systemproblem} introduces the systems of interest, the stability notions and the problems we address. The last subsection collects some known results on the stability of sets for the semigroup generators, as well as an instrumental lemma. Section \ref{Sec:GES} presents the standard, stronger set of control assumptions including invariance and presents both average and measurement dependent switching strategies, for which exponential stability can be proven. We here show that switching based on both fixed dwell-time and hysteresis can be effectively employed.
  Section \ref{Sec:Modulate} relaxes the assumptions, and introduces the new control strategy with the possibility of modulating the controlled dynamics amplitude. In this setting, we can show that the target can be made \emph{globally asymptotically stable} (GAS) even if not all the generators leave it invariant.
  Lastly, Section \ref{Sec:examples} presents numerical simulations that illustrate the effectiveness of the proposed strategies, followed by some comments and outlook in the conclusions.

\textit{Notations:}
The imaginary unit is denoted by $i$.
%$\mathcal{K}$ denotes the set of all continuous functions from $\mathbb{R}_{+}$ to $\mathbb{R}_{+}$ which is strictly increasing and vanishing at zero.
We denote by $\mathcal{B}(\mathcal{H})$ the set of all linear operators on a finite dimensional Hilbert space $\mathcal{H}$. We take $\mathbf{I}$ as the identity operator on $\mathcal{H}$,
%and $\mathbf{I}_R$ as the identity operator on the subspace $\mathcal{H}_R\subset\mathcal{H}$, 
and take $\mathds{1}$ as the indicator function. We denote the adjoint $A\in\mathcal{B}(\mathcal{H})$ by $A^*$. We denote by $\bar{\lambda}(A)$ and $\underline{\lambda}(A)$ the maximum and minimum eigenvalue of the Hermitian matrix $A$ respectively. 
  Define $\mathcal{B}_{*}(\mathcal{H}):=\{X\in\mathcal{B}(\mathcal{H})|X=X^*\}$, $\mathcal{B}_{\geq 0}(\mathcal{H}):=\{X\in\mathcal{B}(\mathcal{H})|X\geq 0\}$ and $\mathcal{B}_{>0}(\mathcal{H}):=\{X\in\mathcal{B}(\mathcal{H})|X> 0\}$, $a\wedge b=\min\{a,b\}$ and $a\vee b=\max\{a,b\}$ for any $a,b\in\mathcal{R}$. 
The function $\mathrm{Tr}(A)$ corresponds to the trace of $A\in\mathcal{B}(\mathcal{H})$. 
%The Hilbert-Schmidt norm of $A\in\mathcal{B}(\mathcal{H})$ is denoted by $\|A\|_{HS}:=\mathrm{Tr}(AA^*)^{1/2}$.
The commutator of two operators $A,B\in\mathcal{B}(\mathcal{H})$ is denoted by $[A,B]:=AB-BA.$ We denote by $\mathrm{int}(\mathcal{S})$ the interior of a subset of a topological space.
\section{System description and problem setting}
\label{sec:systemproblem}
\subsection{Quantum stochastic master equations} We consider quantum systems described on a finite dimensional Hilbert space $\mathcal{H}$. The state of the system is associated to a density matrix on $\mathcal{H}$,  
$$
\mathcal{S}(\mathcal{H}):=\{\rho\in\mathcal{B}(\mathcal{H})|\,\rho=\rho^*\geq 0,\mathrm{Tr}(\rho)=1\}.
$$
Assume that the system is monitored continuously via homodyne/heterodyne measurements, which yield a diffusion observation process. In the quantum filtering regime~\cite{belavkin1989nondemolition,bouten2007introduction,van2007filtering}, the measurements record $Y(t)$ can be described by a continuous semimartingale with quadratic variation $\left\langle Y(t),Y(t)\right\rangle=t$ (see~\cite[Corollary 5.2.9]{van2007filtering} for the proof). Let $\mathcal{F}^Y_t:=\sigma\{Y(s):\, 0\leq s\leq t\}$ be the filteration generated by the observation up to time $t$. The observation process satisfies
\begin{equation*}
dY(t)=\sqrt{\eta}\Tr((C+C^*)\rho(t))dt+dW(t),
\end{equation*} 
where the innovation process $W(t)$ is a one-dimensional Wiener process and satisfies \cite[Proposition 5.2.14]{van2007filtering}  $\mathcal{F}^Y_t=\mathcal{F}_t:=\sigma\{W(s):\, 0\leq s\leq t\}$, $\eta\in(0,1]$ describes the efficiency of the detector, and $C\in\mathcal{B}(\mathcal{H})$ is the noise operator induced by interactions with the probe. 
Conditioning on the observation process $Y(t)$, we have that the conditional density matrix of the system evolves according to a stochastic master equation (SME):
\begin{equation}
d\rho(t)=(-i[H_0,\rho(t)]+\mathcal{D}_C(\rho(t)))dt+\mathcal{G}_C(\rho(t))dW(t),
\label{Eq:SME}
\end{equation}
where $\rho_0\in \mathcal{S}(\mathcal{H})$, $H_0\in\mathcal{B}_*(\mathcal{H})$ is the effective Hamiltonian, which is equal to the free Hamiltonian $H_S$ of the system plus a correction $H_C$ induced by the coupling to the probe, $\mathcal{G}_{C}(\rho):=\sqrt{\eta}(C \rho+\rho C^*-\mathrm{Tr}((C+C^*)\rho)\rho)$, and $\mathcal{D}_C(\rho):=C\rho C^*-C^*C \rho/2-\rho C^*C /2$ is the Lindblad generator associated to the noise operator $C\in\mathcal{B}(\mathcal{H})$. The existence and uniqueness of the solution of the stochastic master equation~\eqref{Eq:SME} as well as the strong Markov property of the solution are ensured by the results established in~\cite{mirrahimi2007stabilizing} or~\cite[Chapter 5]{barchielli2009quantum}. Thanks to the generator $\mathcal{D}_C(\rho)$, $\mathcal{S}(\mathcal{H})$ is almost surely invariant for the stochastic master equation~\eqref{Eq:SME}~\cite[Lemma 3.2, Proposition 3.3]{mirrahimi2007stabilizing}. 

Now, we suppose that the monitored system~\eqref{Eq:SME} can be coupled to one of a finite set of  external systems during assigned periods of time. The effect of these couplings on the dynamics  are our control resources, and which is active at which time is going to be determined by a switching law. Assuming that these external systems act as memory-less (Markov) environments~\cite{alicki2007quantum}, the time evolution of the system state is described by the following switched stochastic master equation,
\begin{equation}
d\rho(t)=\sum^m_{k=1}u^k_t\mathcal{L}_k(\rho(t))dt+\mathcal{G}_{C}(\rho(t))dW(t),
\label{Eq:SSME}
\end{equation}
where $\rho_0\in \mathcal{S}(\mathcal{H})$, for all $k\in\mathcal{M}:=\{1,\dots,m\}$, $\mathcal{L}_k(\rho):=-i[H_0+H_k,\rho]+\mathcal{D}_{L_k}(\rho)+\mathcal{D}_C(\rho)$ with $H_k\in\mathcal{B}_*(\mathcal{H})$  and $L_k\in\mathcal{B}(\mathcal{H})$  representing the Hamiltonian perturbation and the noise operator induced by interactions with the $k$-th external system, respectively, and $u^k_t\in\{0,1\}$ represents the switching law, which is a stochastic process adapted to $\mathcal{F}_t$. This will be designed so that is satisfies, for almost each sample path and $t\geq0$, $u^{i}_t u^{j}_t=0$ for any $i\neq j$ and $\sum^m_{k=1}u^{k}_t=1$, namely only one external system is coupled to the target one at any given time.
The drift term and the diffusion term of the switched stochastic master equation~\eqref{Eq:SSME} satisfy a global random Lipschitz condition. The existence and uniqueness of a global strong solution of the switched system~\eqref{Eq:SSME} 
can be shown by combining the arguments as in~\cite[Theorem 1.2]{krylov1999kolmogorov} and~\cite[Proposition 3.3]{mirrahimi2007stabilizing} or~\cite[Chapter 5]{barchielli2009quantum}. 
\subsection{Invariant and stable subspaces} 
Let $\mathcal{H}_S\subset \mathcal{H}$ be the target subspace. Denote by $\Pi_{\mathfrak{S}}\notin\{0,\mathbf{I}\}$ the orthogonal projection on $\mathcal{H}_S\subset \mathcal{H}$. Define the set of density matrices
\begin{equation*}
\mathcal{I}(\mathcal{H}_S):=\{\rho\in\mathcal{S}(\mathcal{H})| \mathrm{Tr}(\Pi_{\mathfrak{S}}\rho)=1\},
\end{equation*}
namely those whose support is contained in $\mathcal{H}_S$.
\begin{definition}
For the switched system~\eqref{Eq:SSME}, the subspace $\mathcal{H}_S$ is called invariant 
\begin{itemize}
\item in mean if $\rho_0\in \mathcal{I}(\mathcal{H}_S)$, $\mathbb{E}(\rho(t))\in \mathcal{I}(\mathcal{H}_S)$ for all $t>0$.
\item almost surely if $\rho_0\in \mathcal{I}(\mathcal{H}_S)$, $\rho(t)\in \mathcal{I}(\mathcal{H}_S)$ for all $t>0$ almost surely.
\end{itemize}
\end{definition}

Based on the stochastic stability defined in~\cite{khasminskii2011stochastic,mao2007stochastic} and the definition used in~\cite{ticozzi2008quantum,benoist2017exponential}, we phrase the following definition on the stochastic stability of the invariant subspace for the switched system~\eqref{Eq:SSME}. In the following definition, $\|\cdot\|$ could be any matrix norm.

\begin{definition}
\label{def:distance}
Let $\mathcal{H}_S\subset\mathcal{H}$ be the invariant subspace for the switched system~\eqref{Eq:SSME}, and denote by $\Pi_{\mathfrak{S}}$ the orthogonal projection on $\mathcal{H}_S$, $\mathbf{d}_{\mathfrak{S}}(\rho):=\|\rho-\Pi_{\mathfrak{S}}\rho\Pi_{\mathfrak{S}}\|$ and $\hat{\rho}(t):=\mathbb{E}(\rho(t))$, then $\mathcal{H}_S$ is said to be
\begin{enumerate}

\item
\emph{stable in mean}, if there exists class-$\mathcal{K}$ function $\gamma$ such that,
\begin{equation*}
\mathbf{d}_{\mathfrak{S}}(\hat{\rho}(t))\leq \gamma\big(\mathbf{d}_{\mathfrak{S}}(\rho_0)\big) \text{ for } t \geq 0,
\end{equation*}
for all $\rho_0\in \mathcal{S}(\mathcal{H})\setminus \mathcal{I}(\mathcal{H}_S)$.  

\item
\emph{globally asymptotically stable (GAS) in mean}, if it is stable in mean and,
\begin{equation*}
\lim_{t\rightarrow\infty}\mathbf{d}_{\mathfrak{S}}(\hat{\rho}(t))=0, \quad \forall \rho_0\in\mathcal{S}(\mathcal{H}).
\end{equation*}

\item
\emph{globally exponentially stable (GES) in mean}, if there exist two positive constants $\lambda$ and $c$ such that
\begin{equation*}
\mathbf{d}_{\mathfrak{S}}(\hat{\rho}(t))\leq c \,\mathbf{d}_{\mathfrak{S}}(\rho_0) e^{-\lambda t},\quad \forall \rho_0\in\mathcal{S}(\mathcal{H}),
\end{equation*}
where $\lambda$ is called the \emph{average Lyapunov exponent}.

\item
\emph{stable in probability}, if for every $\varepsilon \in (0,1)$, there exists a class-$\mathcal{K}$ function $\gamma$ such that,
\begin{equation*}
\mathbb{P} \left( \mathbf{d}_{\mathfrak{S}}(\rho(t))\leq \gamma\big( \mathbf{d}_{\mathfrak{S}}(\rho_0) \big) \text{ for } t \geq 0 \right) \geq 1-\varepsilon,
\end{equation*}
for all $\rho_0\in \mathcal{S}(\mathcal{H})\setminus \mathcal{I}(\mathcal{H}_S)$. 

\item
\emph{globally asymptotically stable (GAS) almost surely}, if it is stable in probability and,
\begin{equation*}
\mathbb{P} \left( \lim_{t\rightarrow\infty}\mathbf{d}_{\mathfrak{S}}(\rho(t))=0 \right) = 1, \quad \forall \rho_0\in\mathcal{S}(\mathcal{H}).
\end{equation*}

\item
\emph{globally exponentially stable (GES) almost surely}, if
\begin{equation*}
\limsup_{t \rightarrow \infty} \frac{1}{t} \log \big( \mathbf{d}_{\mathfrak{S}}(\rho(t))\big) < 0,\quad \forall \rho_0\in\mathcal{S}(\mathcal{H}), \quad a.s.
\end{equation*}
The left-hand side of the above inequality is called the \emph{sample Lyapunov exponent}.
\end{enumerate}
\end{definition}

The control problem we will be concerned with is the following:\\
\textbf{Switching stochastic stabilization of subspace:}
Given a target subspace $\mathcal{H}_S\subset \mathcal{H}$ and a finite set of generators $\{\mathcal{L}_k(\rho)\}_{k\in\mathcal{M}}$, construct switching laws $u_t$ associated to non-Zeno switching sequence, which ensures that $\mathcal{H}_S$ is GAS and/or GES in mean and almost surely.

\subsection{Stability for semigroup dynamics }
In this section, we recall the invariance and GAS properties of the (Lindblad-Gorini-Kossakowski-Sudarshan) master equation,
\begin{equation}
\frac{d}{dt}\hat{\rho}(t)=\mathcal{L}(\hat{\rho}(t)),\quad \rho_0\in\mathcal{S}(\mathcal{H}),
\label{Eq:MME}
\end{equation}
where $\mathcal{L}(\rho):=i[H,\rho]+\mathcal{D}_L(\rho)+\mathcal{D}_C(\rho)$ is the Lindblad generator, and the unique solution is $\{e^{t\mathcal{L}}\}_{t\geq 0}$ which consists of completely positive trace-preserving map~\cite{ticozzi2008quantum}. This equation corresponds to the semigroup dynamics associated to the expectation of a time-invariant SME of the form \eqref{Eq:SSME}.
Let $\mathcal{H}=\mathcal{H}_S\oplus\mathcal{H}_R$ and $X\in\mathcal{B}(\mathcal{H})$, the matrix representation in an appropriately chosen basis can be written as 
\begin{equation*}
X=\left[\begin{matrix}
X_S & X_P\\
X_Q & X_R
\end{matrix}\right],
\end{equation*}
where $X_S,X_R,X_P$ and $X_Q$ are matrices representing operators from $\mathcal{H}_S$ to $\mathcal{H}_S$, from $\mathcal{H}_R$ to $\mathcal{H}_R$, from $\mathcal{H}_R$ to $\mathcal{H}_S$, from $\mathcal{H}_S$ to $\mathcal{H}_R$, respectively. We denote by $\Pi_{\mathfrak{R}}$ the orthogonal projection on $\mathcal{H}_R\subset\mathcal{H}$. The invariance and GAS properties of the master equation~\eqref{Eq:MME} correspond directly to the structure of Lindblad generator. For the reader's convenience, we summarize some useful results   found in~\cite{ticozzi2008quantum,ticozzi2009analysis,benoist2017exponential}.  
\begin{theorem}
For the system~\eqref{Eq:MME}, the subspace $\mathcal{H}_S$ is 
\begin{enumerate}
\item invariant if and only if $L_Q=C_Q=0$ and $iH_P-\frac{1}{2}(L^*_SL_P+C^*_SC_P)=0$;
\item GAS if and only if it is invariant and no invariant subspaces are included in $\mathrm{ker}(L_{P})\cap\mathrm{ker}(C_{P})$.
\end{enumerate}
\label{Thm:InvarinceGAS}
\end{theorem}
We define the following map,
\begin{equation*}
\mathcal{L}_{R}(\rho_R):=-i[H_{R},\rho_R]+\mathsf{D}_{L}(\rho_R)+\mathsf{D}_{C}(\rho_R),
\end{equation*}
%\begin{equation*}
%\begin{split}
%\mathcal{L}_{j,S}(\rho_S)=-i[H_{j,S},\rho_S]+&\mathcal{D}_{L_{j,S}}(\rho_S)+\mathcal{D}_{C_{S}}(\rho_S),\\
%\mathcal{L}_{j,R}(\rho_R)=-i[H_{j,R},\rho_R]+&L_{j,R}\rho_R L^*_{j,R}-\frac{1}{2}\{L^*_{j,P}L_{j,P}+L^*_{j,R}L_{j,R},\rho_R\}\\
%+&C_{j,R}\rho_R C^*_{j,R}-\frac{1}{2}\{C^*_{j,P}C_{j,P}+C^*_{j,R}C_{j,R},\rho_R\},
%\end{split}
%\end{equation*}
where $\mathsf{D}_{A}(\rho_R):=A_{R}\rho_R A^*_{R}-\frac{1}{2}\{A^*_{P}A_{P}+A^*_{R}A_{R},\rho_R\}$ and $\rho_R\in\mathcal{B}_{\geq 0}(\mathcal{H}_R)$. 
\begin{lemma}
Suppose that $\mathcal{H}_S$ is invariant with respect to the system~\eqref{Eq:MME}. The family $\{e^{t\mathcal{L}_{R}}\}_{t\geq 0}$ is a semigroup of trace non-increasing completely positive maps. Moreover, for all $\rho_0\in\mathcal{S}(\mathcal{H})$, $\hat{\rho}_R(t)=e^{t\mathcal{L}_R}\rho_{0,R}.$
\label{Lemma:LR_TrNonIncreasing}
\end{lemma}
For $R$-block of any Lindblad generator $\mathcal{L}$, we denote by $\mathcal{L}^*_{R}$ the adjoint of $\mathcal{L}_{R}$ with respect to the Hilbert-Schmidt inner product on $\mathcal{B}(\mathcal{H}_R)$. we recall the following results on quantum Markovian dynamics~\cite{benoist2017exponential,evans1977spectral} concerning on the spectral property of $\mathcal{L}^*_R$ and the relation with the GAS in mean with respect to the Lindblad generator $\mathcal{L}$. 
\begin{theorem}
Suppose that $\mathcal{H}_S$ is invariant with respect to the system~\eqref{Eq:MME}. For any $\epsilon>0$, there exists $K_R\in\mathcal{B}_{>0}(\mathcal{H}_R)$ such that 
$
\mathcal{L}_R^*(K_R)\leq -(\alpha-\epsilon)K_R,
$
where $\alpha$ is the spectral abscissa of $\mathcal{L}_R$, i.e., $\alpha:=\min\{-\mathbf{Re}(\lambda)|\,\lambda\in\mathrm{sp}(\mathcal{L}_{R})\}$. Moreover, $\mathcal{H}_S$ is GAS for the system~\eqref{Eq:MME} if and only if $\alpha>0$.
\label{Thm:SpectralAbscissa}
\end{theorem} 

\medskip

Based on the block-decomposition with respect to the orthogonal direct sum decomposition $\mathcal{H}=\mathcal{H}_S\oplus\mathcal{H}_R$, for any  $X_R\in\mathcal{B}(\mathcal{H}_R)$, we call the following matrix the {\em extension of $X_R$ to} $\mathcal{B}(\mathcal{H})$,
\begin{equation*}
X=\left[\begin{matrix}
0 & 0\\
0 & X_R
\end{matrix}\right]\in\mathcal{B}(\mathcal{H}).
\end{equation*} 
In order to quantify the distance between $\rho\in\mathcal{S}(\mathcal{H})$ and $\mathcal{I}(\mathcal{H}_S)$ we shall make use of linear functions associated to a  positive $X_R\in\mathcal{B}_{>0}(\mathcal{H}_R)$, namely
$$
\mathrm{Tr}(X \rho)=\mathrm{Tr}(X_R \rho_R)\in[0,1],
$$
where $X$ is the extension in $\mathcal{B}(\mathcal{H})$ of $X_R$. Such a function is used as an estimation of the distance $\mathbf{d}_{\mathfrak{S}}(\rho)$. 
\begin{lemma}
For all $\rho\in\mathcal{S}(\mathcal{H})$ and the orthogonal projection $\Pi_{\mathfrak{S}}\in\mathcal{B}(\mathcal{H})$ on $\mathcal{H}_S$, there exist two constants $c_1>0$ and $c_2>0$ such that 
\begin{equation}
c_1\mathrm{Tr}(X \rho)\leq \|\rho-\Pi_{\mathfrak{S}}\rho \Pi_{\mathfrak{S}}\| \leq c_2 \sqrt{\mathrm{Tr}(X \rho)},
\label{Eq:Relation_DisLya}
\end{equation}
where $X$ is the extension in $\mathcal{B}(\mathcal{H})$ of $X_R$.
\label{Lemma:Relation_DisLya}
\end{lemma}
\emph{Proof.}
Firstly, let us consider the special case $\mathrm{Tr}(\Pi_{\mathfrak{R}} \rho)=\mathrm{Tr}(\rho_R)$. By employing the arguments in the proof of~\cite[Lemma 4.8]{benoist2017exponential}, we have 
\begin{equation*}
\|\rho-\Pi_{\mathfrak{S}} \rho \Pi_{\mathfrak{S}}\|_{1}\leq N\|\rho-\Pi_{\mathfrak{S}} \rho \Pi_{\mathfrak{S}}\|_{\max}\leq 3N \sqrt{\mathrm{Tr}(\Pi_{\mathfrak{R}} \rho)}
\end{equation*}
where $N=\mathrm{dim}(\mathcal{H})$, $\|\cdot\|_{1}$ and $\|\cdot\|_{\max}$ represents the trace norm and the max norm respectively. Moreover, since the trace norm is unitarily invariant, we have the following pinching inequality~\cite[Chapter 4.2]{bhatia2013matrix}
\begin{align*}
&\|\rho-\Pi_{\mathfrak{S}} \rho \Pi_{\mathfrak{S}}\|_{1}\\
&\geq\| \Pi_{\mathfrak{S}} (\rho-\Pi_{\mathfrak{S}} \rho \Pi_{\mathfrak{S}})\Pi_{\mathfrak{S}}+\Pi_{\mathfrak{R}}(\rho-\Pi_{\mathfrak{S}} \rho \Pi_{\mathfrak{S}})\Pi_{\mathfrak{R}}\|_{1}\\
&=\|\Pi_{\mathfrak{R}}\rho\Pi_{\mathfrak{R}}\|_{1}=\mathrm{Tr}(\Pi_{\mathfrak{R}} \rho).
\end{align*}
Then, the equivalence of matrix norms on finite dimensional vector spaces and the equivalence of $\mathrm{Tr}(\rho_R)$ and $\mathrm{Tr}(X_R \rho_R)$ conclude the proof.
\hfill$\square$

\section{Exponential stabilization of the target subspace}
\label{Sec:GES}

For practical implementations, it is crucial to avoid chattering and Zeno-type phenomena~\cite[Chapter 1.2]{liberzon2003switching} in the switching design. Hysteresis switching technique and fixed dwell time are effective ways to prevent such undesirable behaviors. In the following, we show switching algorithms ensuring GES of the target subspace $\mathcal{H}_S$ under the hysteresis switching and dwell time technique. We start algorithms whose switching sequence can be computed off-line, being based only on the average state evolution which does not depend on the measurement outcomes. 

\subsection{Switching strategy based on the average state}
\label{Subsec:Average}
Here, we reconsider the switching algorithm and control assumptions proposed in~\cite[Theorem 2]{scaramuzza2015switching} for the average dynamics, and show that the same approach leads to GES in mean and almost surely of $\mathcal{H}_S$ for the switching SME~\eqref{Eq:SSME}. The strategy is based on the following invariance and exponential Lyapunov-like assumptions:
\begin{description}
\item[\textbf{A1.1}:] $\mathcal{H}_S$ is invariant with respect to $\mathcal{L}_{k}$ for all $k\in\mathcal{M}$.
\item[\textbf{A1.2}:] 
There exist $K_R\in\mathcal{B}_{>0}(\mathcal{H}_R)$ and $c>0$ such that 
for all $\rho\in\mathcal{S}(\mathcal{H})$,
%for all $\rho\in\mathcal{S}(\mathcal{H})\setminus\mathcal{I}(\mathcal{H_S})$,
$\mathbf{L}_K(\rho)\leq -c\mathrm{Tr}(K\rho)$, where $K$ is the extension in $\mathcal{B}(\mathcal{H})$ of $K_R$ and \[\mathbf{L}_K(\rho):=\min_{k\in\mathcal{M}}\mathrm{Tr}(K\mathcal{L}_k(\rho)).\]
\end{description}
% {\bf [WE HAVE TO SAY HOW IT RELATES TO THE ORIGINAL ASSUMPTION - with convex combination]}
The second assumption is a generalization of the typical assumption for switching stabilizing linear systems, namely the existence of a convex combination of generators that is stabilizing, as it will be made explicit in Corollary \ref{Cor:GES_convex} below. Now, we formulate the average state dependent switching law inspired by~\cite{scaramuzza2015switching} and~\cite[Chapter 3.4]{sun2006switched}. Suppose that \textbf{A1} holds.
For all $j\in\mathcal{M}$, we define the region
\begin{equation}
\Delta_j:=\{\rho\in\mathcal{S}(\mathcal{H})|\,\Tr(K\mathcal{L}_j(\rho))\leq -\epsilon c \Tr(K\rho)\},  
\label{Eq:Region_1}
\end{equation}
where the constant $\epsilon\in(0,1)$ are used to control the dwell-time and the lower bound of the convergence rate. Then, we have 
$
\bigcup_{j\in\mathcal{M}}\mathrm{int}(\Delta_j)=\mathcal{S}(\mathcal{H})\setminus\mathcal{I}(\mathcal{H}_S).
$
Otherwise, there exists a $\rho\in\mathcal{S}(\mathcal{H})\setminus\mathcal{I}(\mathcal{H}_S)$ such that $\Tr(K\mathcal{L}_j(\rho))\geq -\epsilon c \Tr(K\rho)> -c \Tr(K\rho)$ for all $j\in\mathcal{M}$, which leads to a contradiction. Then, we define the switching algorithm $\sigma_1$ based on the average state $\hat{\rho}$.
\begin{definition}[Switching algorithm $\sigma_1$]
For any initial state $\rho_0\in\mathcal{S}(\mathcal{H})\setminus \mathcal{I}(\mathcal{H}_S)$, set $t_0=0$ and for all $n\in\mathbb{N}$,
\begin{align*}
&p_{n}=\textstyle\operatorname*{arg\,min}_{k\in\mathcal{M}}\Tr\big(K\mathcal{L}_k(\hat{\rho}(t_n))\big),\\
&u^{p_{n}}_{t_{n}}=1 \text{ and } u^{k}_{t_{n}}=0, \quad \forall\, k\neq p_{n},\\
&t_{n+1}=\inf\{t\geq t_n|\, \hat{\rho}(t)\notin \mathrm{int}(\Delta_{p_n})\},\\
&u_t \equiv u_{t_n}, \quad \forall\, t\in[t_n,t_{n+1}).
\end{align*}
where throughout this paper we set $\inf\{\emptyset\}=\infty$, and if several Lindbladians correspond to the same decrease rate $p,$ we choose the one with the minimum index. 
%\begin{align*}
%&p_0=\arg\min_{k\in\mathcal{M}}\Tr(K\mathcal{L}_k(\rho_0)), \\
%&u^{p_0}_0=1 \text{ and } u^{k}_0=0, \quad \forall\, k\neq p_0, 
%\end{align*}
%if several Lindbladians yield the same $p_0$, we choose the one with the minimum index. 
%Then, the switching laws $u_t$ and the switching sequence are defined recursively by,
%\begin{align*}
%&t_{n+1}=\inf\{t\geq t_n|\, e^{(t-t_n)\mathcal{L}_{p_n}}\dots e^{t_1\mathcal{L}_{p_0}}\rho_0\notin \mathrm{int}(\Delta_{p_n})\},\\
%&p_{n+1}=\arg\min_{k\in\mathcal{M}}\Tr\big(K\mathcal{L}_k\big(e^{(t_{n+1}-t_n)\mathcal{L}_{p_n}}\dots %e^{t_1\mathcal{L}_{p_0}}\rho_0\big)\big),\\
%&u^{p_{n+1}}_{t_{n+1}}=1 \text{ and } u^{k}_{t_{n+1}}=0, \quad \forall\, k\neq p_{n+1}, 
%\end{align*}
%where throughout this paper we set $\inf\{\emptyset\}=\infty$.
\end{definition}

Note that the above constructed switching instants $\{t_n\}$, switching laws $u_t$ are non-random, and $t_{n+1}>t_n$ since the overlap of each adjacent open regions. Denote by $\bar{n}$ by the total number of switches, which may be infinity. In the following theorem, we show that $\mathcal{H}_S$ is GES in mean and almost surely under the switching algorithm above. Moreover, we provide a lower bound of the dwell time of switching sequence.
Before stating the result, for any $K_R\in\mathcal{B}_{>0}(\mathcal{H}_R)$ and $j\in \mathcal{M}$, we define the following,
\begin{align*}
&\underline{l}_j(K_R):=\sup\{\lambda\in\mathbb{R}|\mathcal{L}^*_{j,R}(K_R)\geq \lambda K_R\}, \\
&\bar{l}_j(K_R):=\inf\{\lambda\in\mathbb{R}|\mathcal{L}^*_{j,R}(K_R)\leq \lambda K_R\},\\
&\underline{l}_{j,2}(K_R):=\sup\{\lambda\in\mathbb{R}|{\mathcal{L}^*}^2_{j,R}(K_R)\geq \lambda K_R\},\\
&\bar{l}_{j,2}(K_R):=\inf\{\lambda\in\mathbb{R}|{\mathcal{L}^*}^2_{j,R}(K_R)\leq \lambda K_R\},\\
&l_j(K_R):=\max\{|\bar{l}_j(K_R)|,|\underline{l}_j(K_R)|\},\\
&l_{j,2}(K_R):=\max\{|\bar{l}_{j,2}(K_R)|,|\underline{l}_{j,2}(K_R)|\}.
\end{align*}
\begin{theorem}
Suppose that \emph{\textbf{A1.1}} and \emph{\textbf{A1.2}} hold true. Then, for the switched system~\eqref{Eq:SSME} under the switching algorithm $\sigma_1$, $\mathcal{H}_S$ is GES in mean and almost surely with the Lyapunov exponent less than or equal to $-\epsilon c/2$. 
\label{Thm:GES_average}
\end{theorem}
\emph{Proof.}
For all $\rho_0\in\mathcal{I}(\mathcal{H}_S)$, the results hold trivially because of the invariance property ensured by \textbf{A1.1}. Let us suppose that $\rho_0\in\mathcal{S}(\mathcal{H})\setminus\mathcal{I}(\mathcal{H}_S)$. First, we show that the switching instants are well-defined, i.e., there exists a constant $t_D>0$ such that $t_{n+1}-t_n>t_D$ for all $n$. For an arbitrary $n$ such that $t_{n+1}<\infty$, due to the assumption \textbf{A1.2} and the definition of $\sigma_1$, we have 
\begin{align*}
&\Tr\big(K(\mathcal{L}_{p_n}(\hat{\rho}(t_n))\big)\leq -c \Tr(K\hat{\rho}(t_{n})), \\
&\Tr\big(K\mathcal{L}_k(\hat{\rho}(t_{n+1}))\big)= -\epsilon c\Tr(K\hat{\rho}(t_{n+1})),
\end{align*}
where $\epsilon\in(0,1)$ and $c>0$ is defined in \textbf{A1.2}. 
For all $t\in[t_n,t_{n+1}]$, we define 
$$
g(t):=\Tr\big(K\mathcal{L}_{p_n}(\hat{\rho}(t))\big)+\epsilon c \Tr(K\hat{\rho}(t)).
$$
One deduces $g(t_n)\leq -c(1-\epsilon)\Tr(K\hat{\rho}(t_{n}))<0$ and $g(t_{n+1})=0$. Moreover, we have
\begin{align*}
|\dot{g}(t)|&\leq \big|\mathrm{Tr}\big((K_R\mathcal{L}^2_{p_n,R}(\hat{\rho}(t)))\big|+\epsilon c\big|\mathrm{Tr}\big(K_R\mathcal{L}_{p_n,R}(\hat{\rho}(t))\big)\big|\\
&\leq \big( l_{p_n,2}(K_R)+\epsilon c l_{p_n}(K_R)\big)\mathrm{Tr}(K\hat{\rho}(t)).
\end{align*}
Due to the mean value theorem, there exits $t^*\in[t_n,t_{n+1}]$ such that
\begin{equation*}
\dot{g}(t^*)(t_{n+1}-t_n)=g(t_{n+1})-g(t_n)\geq c(1-\epsilon)\Tr(K\hat{\rho}(t_{n})).
\end{equation*}
By using Lemma~\ref{Lemma:LR_TrNonIncreasing}, we have 
\begin{align*}
\Tr(K\hat{\rho}(t_{n}))&\geq \underline{\lambda}(K_R)\Tr(\hat{\rho}_R(t_{n}))\geq \underline{\lambda}(K_R)\Tr(\hat{\rho}_R(t^*))\\
&\geq \frac{\underline{\lambda}(K_R)}{\bar{\lambda}(K_R)}\Tr(K\hat{\rho}(t^*)),
\end{align*}
where $\bar{\lambda}(K_R)>0$ and $\underline{\lambda}(K_R)>0$ denote the maximum and minimum eigenvalue of $K_R$ respectively.
It implies 
\begin{align*}
t_{n+1}-t_n&\geq \frac{c(1-\epsilon)}{l_{p_n,2}(K_R)+\epsilon c l_{p_n}(K_R)}\frac{\Tr(K\hat{\rho}(t_{n}))}{\Tr(K\hat{\rho}(t^*))}\\
&\geq \frac{c(1-\epsilon)}{l_{p_n,2}(K_R)+\epsilon c l_{p_n}(K_R)}\frac{\underline{\lambda}(K_R)}{\bar{\lambda}(K_R)}>0.
\end{align*}
Therefore, we have a lower bound of the dwell-time of the switching sequence given by   
\begin{equation}
t_D= \min_{k\in\mathcal{M}} \Big\{ \frac{c(1-\epsilon)}{l_{k,2}(K_R)+\epsilon c l_{k}(K_R)} \frac{\underline{\lambda}(K_R)}{\bar{\lambda}(K_R)}\Big\}>0,
\label{Eq:DwellTime}
\end{equation}
which is negatively correlated to the value of $\epsilon$.

Now, we show $\mathcal{H}_S$ is GES in mean. Based on the definition of switching algorithm, we have,
\begin{align*}
\Tr(K\hat{\rho}(t))&=\Tr(K\rho_0)+\sum^{\bar{n}}_{n=0}\int^{t\wedge t_{n+1}}_{t\wedge t_{n}}\Tr\big(K\mathcal{L}_{p_n}(\hat{\rho}(s))\big)ds\\
&\leq \Tr(K\rho_0)-\epsilon c\int^t_0\Tr(K\hat{\rho}(s))ds.
\end{align*}
By using the Gr\"onwall's inequality, it follows $\Tr(K\hat{\rho}(t))\leq \Tr(K\rho_0)e^{-\epsilon ct}$. Combining with the relation~\eqref{Eq:Relation_DisLya}, one deduces that $\mathcal{H}_S$ is GES in mean with average Lyapunov exponent less than or equal to $-\epsilon c/2$. 

Set $Q_t=e^{\epsilon c t}\Tr(K\rho(t))$, for all $t\geq s\geq 0$, we have
\begin{align*}
\mathbb{E}\big(Q_t\big|\mathcal{F}_s\big)-Q_s=\sum^{\bar{n}}_{n=0}&\int^{t\wedge t_{n+1}}_{s\wedge t_{n}}e^{-\epsilon c u}\mathbb{E}\big( \Tr(K\mathcal{L}_{p_n}\rho(u))\\
&~~~~+\epsilon c\Tr(K\rho(u))\big|\mathcal{F}_s\big)du\leq 0. 
\end{align*}
Thus, $Q_t$ is a positive supermartingale. Due to Doob's martingale convergence theorem~\cite[Corollary 2.11]{revuz2013continuous}, $\sup_{t\geq 0}Q_t=R(\omega)$ almost surely, where $R(\omega)$ is a finite random variable. Moreover, due to Lemma~\ref{Lemma:NeverReach}, it is straightforward to show that $\mathbb{P}\big( \mathrm{Tr}(K\rho(t))>0,\,\forall t\geq0\big)=1$ for all $\rho_0\in\mathcal{S}(\mathcal{H})\setminus\mathcal{I}(\mathcal{H}_S)$. Then, we can deduce $\limsup_{t\rightarrow\infty}\frac{1}{t}\log \Tr(K\rho(t))\leq -\epsilon c$ almost surely. Combining with the relation~\eqref{Eq:Relation_DisLya}, $\mathcal{H}_S$ is GES almost surely with sample Lyapunov exponent less than or equal to $-\epsilon c/2$. 
\hfill$\square$
\begin{remark}
The division of the state space does not satisfy the classical hysteresis technique~\cite{morse1992applications}, since the target subset $\mathcal{I}(\mathcal{H}_S)$ is located at the boundary of all region $\Delta_j$. Due to the bounded convergence rate of $\hat{\rho}(t)$ in each region $\Delta_j$ and the invariance properties of the Lindblad generators (Lemma~\ref{Lemma:LR_TrNonIncreasing}), we can provide a lower bound of the dwell time. However, we cannot determine if the total number of switches $\bar{n}$ is finite or infinite.
\end{remark}

Based on lower bound of dwell time $t_D$ defined in Equation~\eqref{Eq:DwellTime}, we recall the following switching algorithm $\sigma_2$ with fixed dwell time, which is defined in~\cite[Theorem 2]{scaramuzza2015switching} 
\begin{definition}[Switching algorithm $\sigma_2$]
For any initial state $\rho_0\in\mathcal{S}(\mathcal{H})\setminus \mathcal{I}(\mathcal{H}_S)$, fix the dwell time $\Delta t\in(0,t_D]$ where $t_D$ is defined in~\eqref{Eq:DwellTime}, set $t_0 = 0$ and for all $n\in \mathbb{N}$,
\begin{align*}
&p_{n}=\textstyle\operatorname*{arg\,min}_{k\in\mathcal{M}}\Tr\big(K\mathcal{L}_k(\hat{\rho}(t_n))\big),\\
&t_{n+1} = t_n+\Delta t,\\
&u^{p_{n}}_{t}=1 \text{ and } u^{k}_{t}=0, \quad \forall\, k\neq p_{n}, \, t\in[t_n,t_{n+1}).
\end{align*}
%define the switching sequence $t_{n+1}-t_n=\Delta t\leq t_D$ for $n\in\mathbb{N}$ with $t_0=0$, where $t_D$ is defined in %Equation~\eqref{Eq:DwellTime}. Set
%\begin{align*}
%&p_{n}=\arg\min_{k\in\mathcal{M}}\Tr\big(K\mathcal{L}_k\big(e^{(t_{n}-t_{n-1})\mathcal{L}_{p_{n-1}}}\dots %e^{t_1\mathcal{L}_{p_0}}\rho_0\big)\big),\\
%&u^{p_{n}}_{t=1 \text{ and } u^{k}_{t=0, \quad \forall\, k\neq p_{n} \text{ and }\forall\, t\in[t_n,t_{n+1}).
%\end{align*}
\end{definition}
\smallskip
\begin{corollary}
Suppose that \emph{\textbf{A1.1}} and \emph{\textbf{A1.2}} hold true. Then, for the switched system~\eqref{Eq:SSME} under the switching algorithm $\sigma_2$, $\mathcal{H}_S$ is GES in mean and almost surely with the Lyapunov exponent less than or equal to $-\epsilon c/2$. 
\label{Cor:GES_Dwell}
\end{corollary}

The last result of this section shows that our assumption \textbf{A1.2} is implied by the existence of a stabilizing convex combination of generators, and hence we can prove GES under this hypothesis, which is typical in switched control scenarios.  
\begin{corollary}
Suppose that \emph{\textbf{A1.1}} holds true and there exists $\gamma\in\{\gamma\in(0,1)^m| \sum_{j\in\mathcal{M}}\gamma_j=1\}$ such that $\mathcal{H}_S$ is GAS in mean for 
$
\frac{d}{dt}\hat{\rho}(t)=\mathcal{L}_{\gamma}(\hat{\rho}(t))=\sum_{j\in\mathcal{M}}\gamma_j\mathcal{L}_{j}(\hat{\rho}(t)),
$
with $\rho_0\in\mathcal{S}(\mathcal{H})$.
Then, for the switched system~\eqref{Eq:SSME} under the switching algorithm $\sigma_1$ or $\sigma_2$, $\mathcal{H}_S$ is GES in mean and almost surely.
\label{Cor:GES_convex}
\end{corollary}
\emph{Proof.}
Due to Theorem~\ref{Thm:SpectralAbscissa}, the spectral abscissa of $\mathcal{L}_{\gamma}$ denoted by $\alpha_{\gamma}$ is strictly positive. For any $\delta\in(0,\alpha_{\gamma})$, there exists $K_R\in\mathcal{B}_{>0}(\mathcal{H}_R)$ such that $\mathcal{L}^*_{\gamma,R}(K_R)\leq-(\alpha_{\gamma}-\delta)K_R$. Then, we have 
$
\min_{j\in\mathcal{M}}\Tr(K\mathcal{L}_j(\rho))\leq \Tr(K\mathcal{L}_{\gamma}(\rho))\leq -(\alpha_{\gamma}-\delta)\Tr(K\rho),
$
where $K$ is the extension in $\mathcal{B}(\mathcal{H})$ of $K_R$.
Then \textbf{A1.2} is satisfied. By applying Theorem~\ref{Thm:GES_average} and Corollary~\ref{Cor:GES_Dwell}, $\mathcal{H}_S$ is GES in mean and almost surely.
\hfill$\square$

\subsection{Measurement-dependent switching strategies}
In subsection~\ref{Subsec:Average}, we provide switching algorithms ensuring GES of the target subspace $\mathcal{H}_S$ based on the average state evolution. However, by doing so, we do not use the information made available from the measurement output to the fullest. Inspired by~\cite[Theorem 2]{grigoletto2021stabilization} and Theorem~\ref{Thm:GES_average}, we propose two state dependent switching algorithms to guarantee GES of the target subspace $\mathcal{H}_S$ in mean and almost surely. While both aim to select the fastest convergence rate, for each realization, the first one operates with a fixed dwell time, as obtained in the previous results, while the second one can reduce the number of switches by introducing the latter as suitably defined stochastic times.

We define the switching algorithm $\sigma_2$ with fixed dwell time $\Delta t\in(0,t_D]$ based on the state $\rho(t)$, where $t_D>0$ is defined in~\eqref{Eq:DwellTime}.
\begin{definition}[Switching algorithm $\sigma_3$]
For any initial state $\rho_0\in\mathcal{S}(\mathcal{H})\setminus \mathcal{I}(\mathcal{H}_S)$, fix the dwell time $\Delta t\in(0,t_D]$ and set $t_0=0$ and for all $n\in\mathbb{N}$,
\begin{align*}
&p_n(\omega)=\textstyle\operatorname*{arg\,min}_{k\in\mathcal{M}}\Tr(K\mathcal{L}_k(\rho(t_n))), \\
&t_{n+1}=t_n+\Delta t,\\
&u^{k}_{t}(\omega)=\mathds{1}_{\{p_n(\omega)=k\}}, \quad \forall\, k\in\mathcal{M},\,t\in[t_n,t_{n+1}).
\end{align*}
%\begin{align*}
%&t_{n+1}=t_n+\Delta t, \\
%&p_n(\omega)=\arg\min_{k\in\mathcal{M}}\Tr(K\mathcal{L}_k(\rho(t_n))), \\
%&u^{k}_t(\omega)=\mathds{1}_{\{p_n(\omega)=k\}}, \quad \forall\, k\in\mathcal{M},\, t\in[t_n,t_{n+1}),
%\end{align*}
%\begin{equation*}
%\begin{split}
%&p_0=\arg\min_{k\in\mathcal{M}}\Tr(K\mathcal{L}_k(\rho_0)), \\
%&u^{p_0}_0=1 \text{ and } u^{k}_0=0, \quad \forall\, k\neq p_0.
%\end{split}
%\end{equation*}
%Then, the solution of the system~\eqref{Eq:SSME} is well defined for $t\in[t_0,t_1]$. Recursively, the switching law at $t_n$ can be defined by, 
where for each $\omega$, if several $p_n(\omega)$ is active, we choose the minimum, and $u_t$ is $\mathcal{F}_{t_{n}}$-adapted for $t\in[t_n,t_{n+1})$.
\end{definition}
\begin{theorem}
Suppose that \emph{\textbf{A1.1}} and \emph{\textbf{A1.2}} hold true. Then, for the switched system~\eqref{Eq:SSME} under the switching algorithm $\sigma_3$, $\mathcal{H}_S$ is GES in mean and almost surely with the Lyapunov exponent less than or equal to $-\epsilon c/2$. 
\label{Thm:GES_dwell}
\end{theorem}
\emph{Proof.}
For any $n\in\mathbb{N}$ and for all $t\in[t_n,t_{n+1})$, by the construction, the switching law $u_t$ is adapted to $\mathcal{F}_{t_n}$ and
\begin{equation*}
\mathbb{E}(\rho(t)|\mathcal{F}_{t_n})-\rho(t_n)=\mathbb{E}\Big( \int^t_{t_n} \sum^m_{k=1}u^k_s\mathcal{L}_k(\rho(s))ds  \Big| \mathcal{F}_{t_n} \Big),\, a.s.
\end{equation*} 
Set $\Lambda^{k}_{n}:=\{\omega\in\Omega|\, p_n(\omega)=k\}$ with $k\in \mathcal{M}$, then $\mathbb{P}(\bigcup_{k\in\mathcal{M}}\Lambda^{k}_{n})=1$ and $\mathbb{P}(\Lambda^{i}_{n}\cap \Lambda^{j}_{n})=0$ for $i\neq j$. Thus, for almost all $\omega \in \Lambda^{k}_{n}$, due to the linearity of $\mathcal{L}_k$, we have, 
\begin{equation*}
\mathbb{E}(\rho(t)|\mathcal{F}_{t_n})-\rho(t_n)=\int^t_{t_n} \mathcal{L}_k\big( \mathbb{E}(\rho(s)|\mathcal{F}_{t_n}) \big)ds,
\end{equation*}
It follows that, for almost all $\omega \in \Lambda^{k}_{n}$ and $t\in[t_n,t_{n+1})$, $\mathbb{E}(\rho(t)|\mathcal{F}_{t_n})$ solves the master equation generated by $\mathcal{L}_k$. 
By using the similar arguments as in the proof of Theorem~\ref{Thm:GES_average} and Corollary~\ref{Cor:GES_Dwell}, we have  
$$
\Tr\big(K\mathcal{L}_{k}(\mathbb{E}(\rho(t)| \mathcal{F}_{t_n}))\big)\leq -\epsilon c \Tr\big(K\mathbb{E}(\rho(t)| \mathcal{F}_{t_n})\big)
$$
on the set $\Lambda^{k}_{n}$. Since $k$ is chosen arbitrarily, we have  
\begin{align*}
&\mathbb{E}\big(\Tr(K\rho(t)) \big| \mathcal{F}_{t_n} \big)-\Tr(K\rho(t_n))\\
&\leq -\epsilon c \,\mathbb{E}\Big(\int^t_{t_n}\Tr(K\rho(s))ds \Big| \mathcal{F}_{t_n} \Big), \quad a.s. 
\end{align*}
Due to the law of total expectation, we have 
\begin{equation*}
\Tr(K\hat{\rho}(t))-\Tr(K\hat{\rho}(t_n))\leq -\epsilon c \int^t_{t_n}\Tr(K\hat{\rho}(s))ds 
\end{equation*}
Combine each switching interval together, then apply the Gr\"onwall's inequality and the relation~\eqref{Eq:Relation_DisLya}, one deduces that $\mathcal{H}_S$ is GES in mean with average Lyapunov exponent less than or equal to $-\epsilon c/2$. 

Denote $Q_t=e^{\epsilon c t}\Tr(K\rho(t))$. For all $0\leq s \leq t<\infty$, there exist $l,d\in\mathbb{N}$ such that $t_{l}\leq s\leq t_{l+1}\leq \dots \leq t_{l+d} \leq t$. By It\^o formula~\cite[Theorem 2.32]{protter2004stochastic}, we have
\begin{equation*}
\mathbb{E}(Q_t|\mathcal{F}_s)-Q_s=\sum^{l+d}_{j=l}I_s(s\vee t_{j},t\wedge t_{j+1}),\quad a.s.
\end{equation*}
where $I_s(\alpha,\beta):=\mathbb{E}\big(\int^{\beta}_{\alpha} e^{\epsilon c r}\big(\sum^m_{k=1}u^k_r\Tr\big(K\mathcal{L}_k(\rho(r))\big)+\epsilon c \Tr(K\rho(r))  \big)dr   \big|\mathcal{F}_s  \big)$. Based on the above arguments, for almost each $\omega\in\Lambda^k_n$ with $k\in\mathcal{M}$ and $n\in\{l,\dots,l+d\}$, we have
$I_s(\alpha,\beta)\leq 0$. Thus, $Q_t$ is a positive supermartingale. By employing the similar arguments in the proof of Theorem~\ref{Thm:GES_average}, $\mathcal{H}_S$ is GES almost surely with sample Lyapunov exponent less than or equal to $-\epsilon c/2$.
\hfill$\square$

\medskip

Next, we define the switching algorithm $\sigma_4$ based on the state $\rho(t)$.
For any initial state $\rho_0\in\mathcal{S}(\mathcal{H})\setminus \mathcal{I}(\mathcal{H}_S)$, set $\tau_0=0$ and 
\begin{align*}
&p_0=\textstyle\operatorname*{arg\,min}_{k\in\mathcal{M}}\Tr(K\mathcal{L}_k(\rho_0)), \\
&u^{p_0}_0=1 \text{ and } u^{k}_0=0, \quad \forall\, k\neq p_0.
\end{align*}
The solution of the switched stochastic master equation~\eqref{Eq:SSME} under the switching law $u_t\equiv u_0$ is well-defined on $t\in[\tau_0,\infty)$. Define the stopping time 
\begin{equation*}
\tau_1:=\inf\{t\geq \tau_0|\, \rho(t)\notin \mathrm{int}(\Delta_{p_0}) \},
\end{equation*}
where $\Delta_j=\{\rho\in\mathcal{S}(\mathcal{H})|\,\Tr(K\mathcal{L}_j(\rho))\leq -\epsilon c \Tr(K\rho)\}$ with $j\in\mathcal{M}$.
Then, for any finite $T\geq 0$, we define
\begin{equation*}
\Omega^1_T:=\{\omega\in\Omega|\, \tau_1<T\}, \, \bar{\Omega}^1_T:=\{\omega\in\Omega|\, \tau_1\geq T\}.
\end{equation*}
Obviously, we have $\mathbb{P}(\Omega^1_T\cup \bar{\Omega}^1_T)=1$ and $\mathbb{P}(\Omega^1_T\cap \bar{\Omega}^1_T)=0$.
For almost each $\omega \in \bar{\Omega}^1_T$, the trajectory of the system stays in $\mathrm{int}(\Delta_{p_0})$ without switching till $T$. For all $\omega \in \Omega^1_T$, we define
\begin{align*}
&p_1(\omega)=\textstyle\operatorname*{arg\,min}_{k\in\mathcal{M}}\Tr\big(K \mathcal{L}_k (\rho(\tau_1))\big),\\
&u^{k}_t(\omega)=\mathds{1}_{\{p_1(\omega)=k\}}, \quad \forall\, k\in\mathcal{M},\, t\in[\tau_1,T).
\end{align*}
The second switching instant is defined as
\begin{equation*}
    \tau_2(\omega):=\inf\{t\geq \tau_1|\, \rho(t)\notin \mathrm{int}(\Delta_{p_1(\omega)}) \}.
\end{equation*}
We denote
\begin{equation*}
\Omega^2_T:=\{\omega\in\Omega^1_T|\, \tau_2<T\}, \, \bar{\Omega}^2_T:=\{\omega\in\Omega^1_T|\, \tau_2\geq T\},
\end{equation*}
which follows $\mathbb{P}(\Omega^2_T\cup \bar{\Omega}^2_T)=\mathbb{P}(\Omega^1_T)$ and $\mathbb{P}(\Omega^2_T\cap \bar{\Omega}^2_T)=0$.
Then, we can define the switching laws at switching instants $\tau_n$ and $\Omega^n_T$ and $\bar{\Omega}^n_T$ for all $n\in\mathbb{N}$ recursively. 
\begin{definition}[Switching algorithm $\sigma_4$]
For any initial state $\rho_0\in\mathcal{S}(\mathcal{H})\setminus \mathcal{I}(\mathcal{H}_S)$, set $\tau_0=0$ and for all $n\in\mathbb{N}$,
\begin{align*}
&p_{n}(\omega)=\textstyle\operatorname*{arg\,min}_{k\in\mathcal{M}}\Tr\big(K \mathcal{L}_k (\rho(\tau_{n}))\big),\\
&u^{k}_{\tau_n}(\omega)=\mathds{1}_{\{p_{n}(\omega)=k\}}, \quad \forall\, k\in\mathcal{M},\\
&\tau_{n+1}(\omega):=\inf\{t\geq \tau_n|\, \rho(t)\notin \mathrm{int}(\Delta_{p_{n}(\omega)}) \},\\
&u_t(\omega)\equiv u_{\tau_n}(\omega),\quad \forall\, t\in[\tau_n,\tau_{n+1}).
\end{align*}
%\begin{align*}
%&p_{n}(\omega)=\arg\min_{k\in\mathcal{M}}\Tr\big(K \mathcal{L}_k (\rho(\tau_{n}))\big),\\
%&u^{k}_t(\omega)=\mathds{1}_{\{p_{n}(\omega)=k\}}, \quad \forall\, k\in\mathcal{M},\, t\in[\tau_{n}(\omega),\tau_{n+1}(\omega)),\\
%&\tau_{n+1}(\omega):=\inf\{t\geq \tau_n(\omega)|\, \rho(t)\notin \mathrm{int}(\Delta_{p_{n}(\omega)}) \}.
%\end{align*}
\end{definition}
\smallskip

Due to the non-empty overlap of each adjacent open regions $\mathrm{int}(\Delta_j)$, the continuity of the solution $\rho(t)$ for almost each sample path on $t\in[0,T]$, and Lemma~\ref{Lemma:NeverReach} which implies the almost sure inaccessibility of $\mathcal{I}(\mathcal{H}_S)$ in finite time, one deduces that the number of switches before any finite time $T$ is finite almost surely, which is denoted by $\bar{n}_T(\omega)<\infty$.
For any positive constant $T<\infty$, the switching law $(u_t)_{t\in[0,T]}$ is adapted to $\mathcal{F}_T$, the solution of switching solution~\eqref{Eq:SSME} is well-defined.

\begin{theorem}
Suppose that \emph{\textbf{A1.1}} and \emph{\textbf{A1.2}} hold true. Then, for the switched system~\eqref{Eq:SSME} under the switching algorithm $\sigma_4$, $\mathcal{H}_S$ is GES in mean and almost surely with the Lyapunov exponent less than or equal to $-\epsilon c/2$. 
\label{Thm:GES_state}
\end{theorem}
\emph{Proof.}
Fix an arbitrary positive constant $T<\infty$ and $n\in\mathbb{N}$. Suppose $\rho_0\in\mathcal{S}(\mathcal{H})\setminus \mathcal{I}(\mathcal{H}_S)$, for almost each $\omega\in\bar{\Omega}^n_T$, we have
\begin{align*}
&\int^T_0 \sum^m_{k=1} u^k_{s}(\omega) \Tr(K\mathcal{L}_k \rho(s))ds\\
&=\sum^{n-1}_{j=0}\int^{T\wedge \tau_{j+1}(\omega)}_{\tau_{j}(\omega)} \sum^m_{k=1} u^k_{s}(\omega) \Tr(K\mathcal{L}_k \rho(s))ds\\
&\leq -\epsilon c \int^T_0 \Tr(K\rho(s))ds.
\end{align*}
Since $\bar{n}_T<\infty$ almost surely, $\mathbb{P}(\lim_{n\rightarrow \infty} \bigcup^n_{i=1}\bar{\Omega}^i_T) =1$. Due to It\^o isometry, we obtain
\begin{align*}
&\Tr(K\hat{\rho}(T)) -\Tr(K\rho_0)\\
&=\mathbb{E}\Big( \int^T_0 \sum^m_{k=1} u^k_{s}(\omega) \Tr(K\mathcal{L}_k \rho(s))ds  \Big)\\
&=\mathbb{E}\Big( \lim_{n\rightarrow \infty} \sum^n_{i=1}\mathds{1}_{\bar{\Omega}^i_T}\int^T_0 \sum^m_{k=1} u^k_{s}(\omega) \Tr(K\mathcal{L}_k \rho(s))ds  \Big)\\
&\leq -\epsilon c\,\mathbb{E}\Big( \lim_{n\rightarrow \infty} \sum^n_{i=1}\mathds{1}_{\bar{\Omega}^i_T}\int^T_0 \Tr(K\rho(s))ds  \Big)\\
&= -\epsilon c\, \int^T_0 \Tr(K\hat{\rho}(s))ds  
\end{align*}
By applying the Gr\"onwall's inequality, in addition to the relation~\eqref{Eq:Relation_DisLya}, one deduces that $\mathcal{H}_S$ is GES in mean with average Lyapunov exponent less than or equal to $-\epsilon c/2$. 

Denote $Q_t=e^{\epsilon c t}\Tr(K\rho(t))$. For all $0\leq s \leq t<\infty$, by It\^o formula, we have
\begin{equation*}
\mathbb{E}(Q_t|\mathcal{F}_s)-Q_s=I_s(s,t),\quad a.s.
\end{equation*}
where 
\begin{align*}
I_s(s,t):=\mathbb{E}\big(\int^{t}_{s} e^{\epsilon c r}\big(&\sum^m_{k=1}u^k_r\Tr\big(K\mathcal{L}_k(\rho(r))\big)\\
&+\epsilon c \Tr(K\rho(r))  \big)dr   \big|\mathcal{F}_s  \big).
\end{align*}
Based on the above arguments, for almost all $\omega\in\bar{\Omega}^k_t$ with $k\in\mathcal{M}$, we have $I_s(s,t)\leq 0$. Thus, $Q_t$ is a supermartingale since $\mathbb{P}(\lim_{n\rightarrow \infty} \bigcup^n_{i=1}\bar{\Omega}^i_T)=1$. By employing the similar arguments in the proof of Theorem~\ref{Thm:GES_average}, $\mathcal{H}_S$ is GES almost surely with sample Lyapunov exponent less than or equal to $-\epsilon c/2$.
\hfill$\square$

\section{Asymptotic stabilization of target subspace by modulating Lindbladian}
\label{Sec:Modulate}
From the practical point of view, the assumption \textbf{A1.1} and \textbf{A1.2} might be too restrictive. In particular, the assumption \textbf{A1.1} on the invariance property of the target subspace $\mathcal{H}_S$ limits significantly the type of control actions that can be employed. Inspired by~\cite[Theorem 3]{grigoletto2021stabilization}, we relax such assumptions to the following:
\begin{description}
\item[\textbf{A2}:] There exists a $K_R\in\mathcal{B}_{>0}(\mathcal{H}_R)$ such that $\mathbf{L}_K(\rho)<0$ for all $\rho\in\mathcal{S}(\mathcal{H})\setminus\mathcal{I}(\mathcal{H}_S)$, where $K$ is the extension in $\mathcal{B}(\mathcal{H})$ of $K_R$.
\end{description}
Lemma~\ref{Lem:Continuity} shows that $\mathbf{L}_K(\rho)$ is continuous on $\mathcal{S}(\mathcal{H})$. In addition to the compactness of $\Theta_{\delta}:=\{\rho\in\mathcal{S}(\mathcal{H})|\,\Tr(K\rho)\geq \delta\}$ with $\delta>0$ and the assumption \textbf{A2}, there exists a constant $\gamma(\delta)>0$ such that $\mathbf{L}_K(\rho)\leq -\gamma(\delta)$ for all $\rho\in\Theta_{\delta}$. Then, we can deduce the average practical stability of the switched systems~\eqref{Eq:SSME} with a dwell time dependent on $\delta$ under \textbf{A2}. See~\cite[Theorem 3]{grigoletto2021stabilization} for the details. However, due to Lemma~\ref{Lem:Min=0}, $\gamma(\delta)$ converges to zero when $\delta$ tends to zero. Hence, we cannot fix a dwell time such that $\Tr(K\rho(t))$ decreases during each switching interval. 

This issue can be addressed by considering {\em a modulated gain for the active Lindbaldian during each switching interval}, so that the speed of the derivative of $\Tr(K\rho(t))$ increasing to zero can be reduced. This idea can be leveraged to guarantee GAS of the target subspace $\mathcal{H}_S$ in mean and almost surely with an arbitrary non-zero dwell time and without any requirement on the invariance properties of Lindblad generators. Thus, from now on, we suppose that the gains of Lindbladians are adjustable. The dynamics of the switched stochastic master equation is given by, 
\begin{align}
d\rho(t)=\sum^m_{k=1}v^k_t\mathcal{L}_k(\rho(t))dt+\sum^m_{k=1}v^k_t\mathcal{G}_{C}(\rho(t))dW(t),
\label{Eq:SSME_modulate}
\end{align} 
where $\rho_0\in \mathcal{S}(\mathcal{H})$, $v^k_t\in[0,V]$ with $V<\infty$ and $k\in\mathcal{M}$ represents the switching control laws, which is bounded random variable adapted to $\mathcal{F}_t$, and for almost every sample path and $t\geq 0$, $v^{i}_t v^{j}_t=0$ with $i\neq j$. The drift and diffusion term of the switched system~\eqref{Eq:SSME_modulate} still obey a global random Lipschitz condition, so existence and uniqueness of a global strong solution can be proved as before. 

Next, we introduce the switching strategy $\sigma_5$ with the fixed dwell time based on $\rho(t)$. Fix an arbitrary dwell time $\Delta t>0$ and define $t_n=n\Delta t$ for any $n\in\mathbb{N}$. Set
\begin{align*}
&p_n(\omega)=\textstyle\operatorname*{arg\,min}_{k\in\mathcal{M}}\Tr(K\mathcal{L}_k(\rho(t_n))), \\
&v^{k}_t(\omega)=q_n(\omega)\mathds{1}_{\{p_n(\omega)=k\}}, \quad \forall\, k\in\mathcal{M},\, t\in[t_n,t_{n+1}),
\end{align*}
where $q_n$ is a bounded random variable adapted to $\mathcal{F}_{t_n}$. Set $q_n=0$ if $\Tr(K\rho(t_n))=0$. Due to the relation~\eqref{Eq:Relation_DisLya}, $\rho(t)$ stays in $\mathcal{I}(\mathcal{H}_S)$ afterwards in this case. For the case $\Tr(K\rho(t_n))>0$, $q_n$ will be determined later. Then, the solution of the switched system~\eqref{Eq:SSME_modulate} is well-defined on $[0,t_{n+1}]$. By It\^o formula, for $t\in[t_n,t_{n+1}]$, we have
\begin{equation*}
\mathbb{E}(\rho(t)|\mathcal{F}_{t_n})-\rho(t_n)=\mathbb{E}\Big( \int^t_{t_n} \sum^m_{k=1}v^k_s\mathcal{L}_k(\rho(s))ds  \Big| \mathcal{F}_{t_n} \Big), \,a.s.
\end{equation*} 
Set $\Lambda^k_n:=\{\omega\in\Omega|\, p_n(\omega)=k \text{ and }\Tr(K\rho(t_n))>0 \}$ with $k\in \mathcal{M}$, then $\mathbb{P}(\bigcup_{k\in\mathcal{M}}\Lambda^{k}_{n})=\mathbb{P}(\Tr(K\rho(t_n))>0)$ and $\mathbb{P}(\Lambda^{i}_{n}\cap \Lambda^{j}_{n})=0$ for $i\neq j$.
For almost all $\omega\in\Lambda^k_n$, due to the linearity of $\mathcal{L}_k$, we have
\begin{equation*}
\mathbb{E}(\rho(t)|\mathcal{F}_{t_n})-\rho(t_n)=\int^t_{t_n} q_n(\omega)\mathcal{L}_k\big( \mathbb{E}(\rho(s)|\mathcal{F}_{t_n}) \big)ds,
\end{equation*}
where we used the fact that $q_n$ is adapted to $\mathcal{F}_{t_n}$. It implies that, for almost all $\omega\in\Lambda^k_n$, $\mathbb{E}(\rho(t)|\mathcal{F}_{t_n})=e^{(t-t_n)q_n(\omega)\mathcal{L}_k}\rho(t_n)$. Then, for almost every $\omega\in\Lambda^k_n$, 
\begin{align*}
&\mathbb{E}(\Tr(K\mathcal{L}_k(\rho(t)))|\mathcal{F}_{t_n})-\Tr(K\mathcal{L}_k\rho(t_n))\\
&=\mathbb{E}\Big( \int^{t}_{t_n} q_n(\omega)\Tr(K\mathcal{L}^2_k(\rho(s)))ds  \Big| \mathcal{F}_{t_n} \Big)\\
&\leq \bar{M}_K (t-t_n) q_n(\omega),
\end{align*}
where $\bar{M}_K:=\max_{k\in\mathcal{M}}\sup_{\rho\in\mathcal{S}(\mathcal{H})}\Tr(K\mathcal{L}^2_k(\rho))$. One deduces that, for all $t\in[t_n,t_{n+1}]$, 
\begin{align*}
&\mathbb{E}(\Tr(K\mathcal{L}_k(\rho(t)))|\mathcal{F}_{t_n})\\
&=\Tr\big(K\mathcal{L}_k\big(e^{(t-t_n)q_n(\omega)\mathcal{L}_k}\rho(t_n)\big)\big)\\
&\leq \bar{M}_K (t-t_n) q_n(\omega)+\Tr(K\mathcal{L}_k\rho(t_n)),
\end{align*}
where $\Tr(K\mathcal{L}_k\rho(t_n))<0$ due to the assumption \textbf{A2}. Define 
\begin{equation*}
q_n(\omega):=-\frac{\Tr(K\mathcal{L}_k\rho(t_n))}{\bar{M}_K \Delta t }>0, \quad \forall \, \omega\in\Lambda^k_n.
\end{equation*}
Hence, for almost all $\omega\in\Lambda^k_n$ and $t\in[t_n,t_{n+1})$, 
\begin{equation*}
\Tr\big(K\mathcal{L}_k\big(e^{(t-t_n)q_n(\omega)\mathcal{L}_k}\rho(t_n)\big)\big)<0.
\end{equation*}
Then, for almost each $\omega\in\Lambda^k_n$, we have
\begin{align*}
&\mathbb{E}(\Tr(K\rho(t_{n+1}))|\mathcal{F}_{t_n})-\Tr(K\rho(t_n))\\
&=\mathbb{E}\Big( \int^{t_{n+1}}_{t_n} q_n(\omega)\Tr(K\mathcal{L}_k(\rho(s)))ds  \Big| \mathcal{F}_{t_n} \Big)\\
&=\int^{t_{n+1}}_{t_n} q_n(\omega)\Tr\big(K\mathcal{L}_k\big(e^{(s-t_n)q_n(\omega)\mathcal{L}_k}\rho(t_n)\big)\big)ds<0.
\end{align*}
Due to the arbitrariness of $k$, this implies the strict decrease of 
$
\mathbb{E}(\Tr(K\rho(t))|\mathcal{F}_{t_n})
$
on $[t_n,t_{n+1})$ when $\Tr(K\rho(t_n))>0$. We are led to the following:
\begin{definition}[Switching strategy $\sigma_5$]
Fix the dwell time $\Delta t>0$ and set $t_0=0$,
\begin{align*}
&p_n(\omega)=\textstyle\operatorname*{arg\,min}_{k\in\mathcal{M}}\Tr(K\mathcal{L}_k(\rho(t_n))), \\
&q_n(\omega)=
\begin{cases}
-\frac{\Tr(K\mathcal{L}_k\rho(t_n))}{\bar{M}_K \Delta t },& \text{if } \Tr(K\rho(t_n))>0,\\
0,& \text{else},
\end{cases}\\
&t_{n+1}=t_n+\Delta t,\\
&v^{k}_t(\omega)=q_n(\omega)\mathds{1}_{\{p_n(\omega)=k\}}, \quad \forall\, k\in\mathcal{M},\, t\in[t_n,t_{n+1}).
\end{align*}
\end{definition}
\smallskip
\begin{theorem}
Suppose that \emph{\textbf{A2}} holds true. For any finite dwell time $\Delta t>0$, $\mathcal{H}_S$ is GAS in mean and almost surely for the switched system~\eqref{Eq:SSME_modulate} under the switching strategy $\sigma_5$.
\label{Thm:GAS}
\end{theorem}
\emph{Proof.}
If there exists $\bar{n}<\infty$ such that $\Tr(K\rho(t_{\bar{n}}))=0$ almost surely, it reduces to a trivial case. Suppose that, for all $\rho_0\in\mathcal{S}(\mathcal{H})\setminus \mathcal{I}(\mathcal{H}_S)$, $\mathbb{P}(\Tr(K\rho(t_{n})>0,\, \forall n\in\mathbb{N})>0$.
Due to the law of total expectation, for any $n\in\mathbb{N}$,
\begin{align*}
&\mathbb{E}\big( \Tr(K\rho(t_{n+1})) \big)-\mathbb{E}\big( \Tr(K\rho(t_{n})) \big)\\
&=\mathbb{E}\big(\mathbb{E}(\Tr(K\rho(t_{n+1}))|\mathcal{F}_{t_n})-\Tr(K\rho(t_n)) \big)\\
&=\mathbb{E}\big(\mathds{1}_{\{\Tr(K\rho(t_{n})>0\}}\big( \mathbb{E}(\Tr(K\rho(t_{n+1}))|\mathcal{F}_{t_n})\\
&~~~~~~~~~~~~~~~~~~~~~~~~~~~~~~-\Tr(K\rho(t_n)) \big) \big)<0.
\end{align*}
Then, by employing the standard Lyapunov arguments~\cite[Theorem 10.1.1]{hale1980ordinary}, GAS in mean of $\mathcal{H}_S$ can be concluded. Moreover, due to Chebyshev’s inequality, $\mathcal{H}_S$ is stable in probability~\cite[Theorem 5.3]{khasminskii2011stochastic}. By using the dominated convergence theorem, almost sure GAS of $\mathcal{H}_S$ can be concluded.
\hfill$\square$

%\section{Discussion}

\section{Numerical simulations} \label{Sec:examples}
In this section, we illustrate the performance of the proposed switching strategies in stabilizing a three-qubit system and a spin-$\frac{3}{2}$ system towards a pre-determined GHZ state and a pre-determined target space, respectively. In order to ensure that the trajectories of switched stochastic master equation~\eqref{Eq:SSME} stay in $\mathcal{S}(\mathcal{H})$ during the simulation, we employ the numerical scheme proposed in~\cite{rouchon2015efficient}. 
%The following simulations have been run with step length $\delta t = 0.002$, number of steps $N = 5000$, each with $1000$ realizations. 

\subsection{GHZ stabilization under assumption \emph{\textbf{A1}}}
Here, we apply the switching strategies $\sigma_1$, $\sigma_2$, $\sigma_3$ and $\sigma_4$ on the three-qubit system proposed in~\cite{ticozzi2014steady,liang2022GHZ}. { For this system, invariance is satisfied and open-loop stabilization schemes exist: the main role of switching and feedback is to speed-up the convergence. In the following we shall highlight the advantage of the proposed scheme with respect to simpler engineered dissipation methods.}

\begin{figure}[!t]
\centerline{\includegraphics[width=\columnwidth]{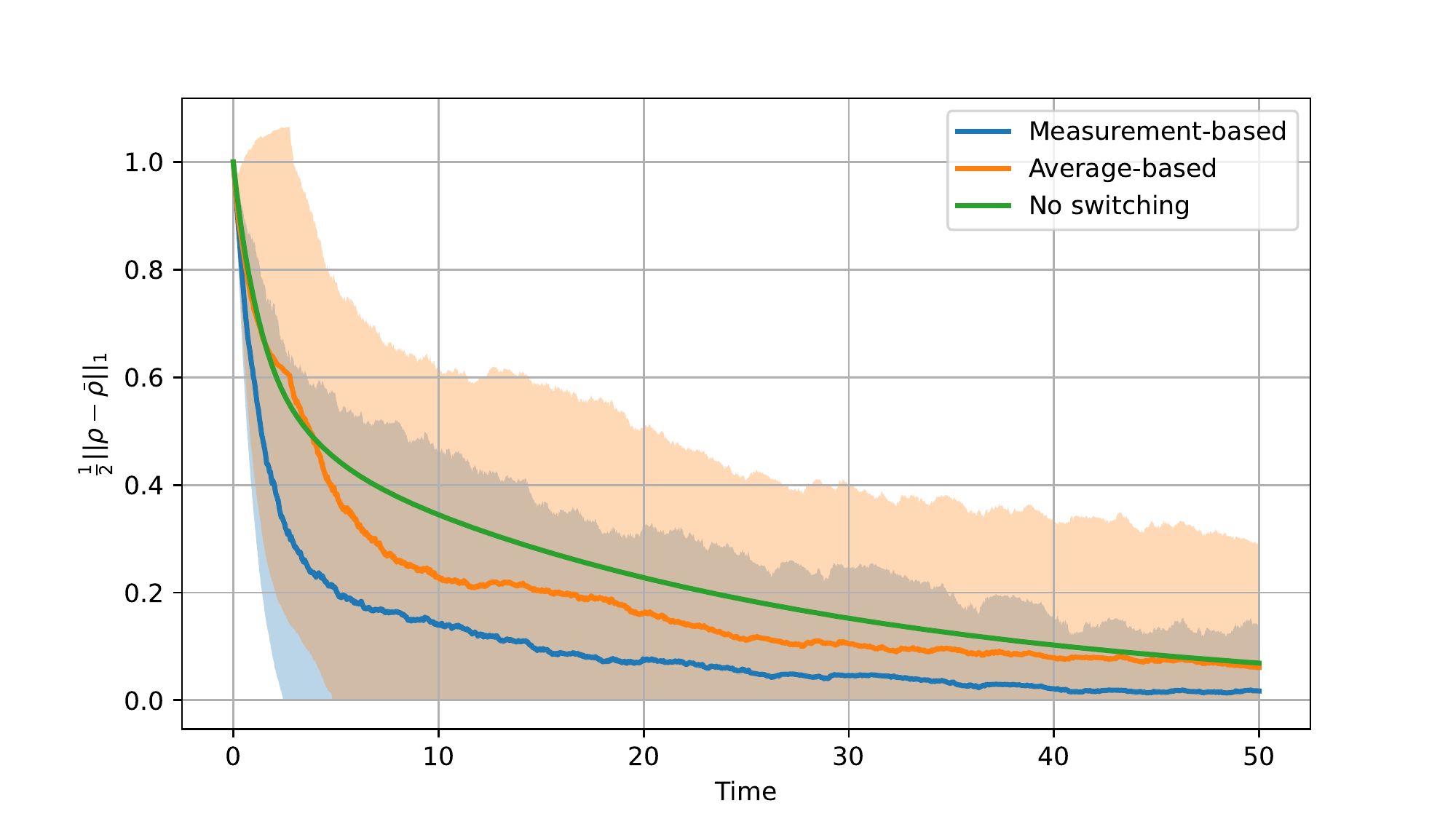}}
\caption{Exponential stabilization of a three-qubit system toward $\bar{\rho}$ starting from $\rho_0$ with fixed dwell time $t_D$. The blue curve represents the mean value of 200 measurement-based {($\sigma_2$)} realizations and the light blue area shows the mean value plus or minus one standard deviation, the orange curve represents the mean value of 200 average-based {($\sigma_3$)} realizations and the light orange area shows the mean value plus or minus one standard deviation. {The green curve represents the evolution given by the generator $\mathcal{L}(\cdot) = \frac{1}{2}[\mathcal{L}_1(\cdot)+\mathcal{L}_2(\cdot)]$.} }
\label{SIM:Fixed}
\end{figure}

\begin{figure}[!t]
\centerline{\includegraphics[width=\columnwidth]{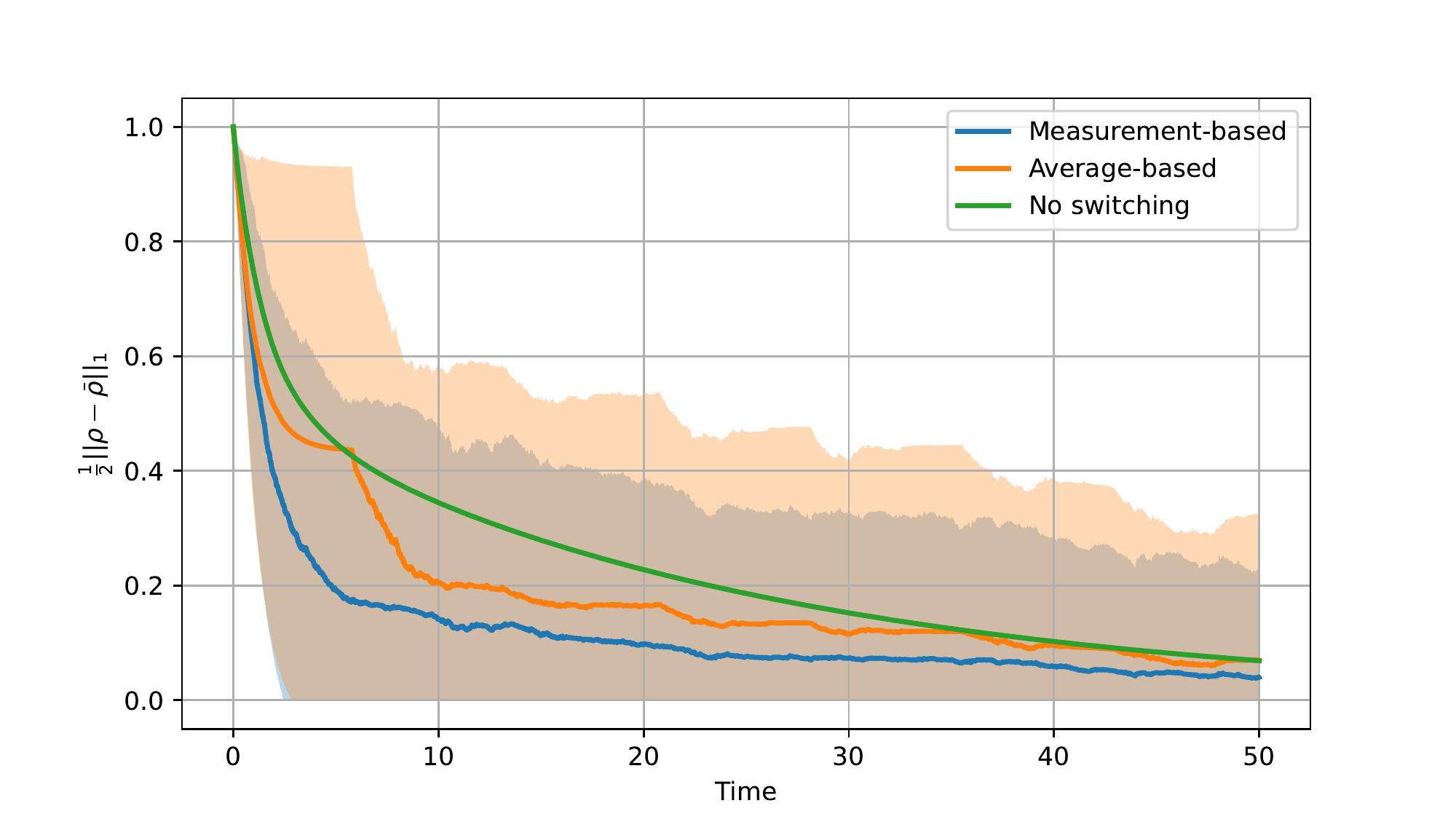}}
\caption{Exponential stabilization of a three-qubit system toward $\bar{\rho}$ starting from $\rho_0$ under hysteresis switching. The blue curve represents the mean value of 200 {measurement-based ($\sigma_4$)} realizations and the light blue area shows the mean value plus or minus one standard deviation, the orange curve represents the mean value of 200 {average-based ($\sigma_1$)} realizations and the light orange area shows the mean value plus or minus one standard deviation. {The green curve represents the evolution given by the generator $\mathcal{L}(\cdot) = \frac{1}{2}[\mathcal{L}_1(\cdot)+\mathcal{L}_2(\cdot)]$.}}
\label{SIM:Hysteresis}
\end{figure}

The target state is determined as 
\begin{equation*}
    \bar{\rho} = \frac{1}{2}(|000\rangle+|111\rangle)(\langle 000|+\langle 111|),
\end{equation*}
and the initial state is set as $\rho_0 = \frac{1}{4}|\Psi\rangle\langle\Psi|$ where
\begin{equation*}
    |\Psi\rangle = |001\rangle+|010\rangle+|101\rangle+|110\rangle.
\end{equation*}
We construct two Lindbladian generators as follows
\begin{align*}
    &H_1 = H_2 = \sigma_x\otimes \mathbf{I}\otimes \mathbf{I}-\mathbf{I}\otimes \sigma_x\otimes\sigma_x;\\
    &L_1 = (|00\rangle\langle 01|+|11\rangle\langle10|)\otimes \mathbf{I};\\
    &L_2 = \mathbf{I}\otimes (|00 \rangle \langle 01|+i|11\rangle\langle10|);\\
    &C = \sigma_z\otimes \mathbf{I}\otimes \sigma_z,
\end{align*}

where $\sigma_x$ and $\sigma_z$ are Pauli matrices. Based on Theorem~\ref{Thm:InvarinceGAS}, we can verify that $\bar{\rho}$ is invariant for the Lindbladian generators { $\mathcal{L}_1(\cdot) = -i[H_1, \cdot] + \mathcal{D}_{L_1}(\cdot) + \mathcal{D}_C(\cdot)$ and $\mathcal{L}_2(\cdot) = -i[H_2, \cdot] + \mathcal{D}_{L_2}(\cdot) + \mathcal{D}_C(\cdot)$,} yet not GAS under any single Lindbladian. However, the simultaneous action of $\mathcal{L}_1$ and $\mathcal{L}_2$, with arbitrary positive weights, leads to the GAS of $\bar{\rho}$. Then, the assumption \textbf{A1} is satisfied due to Corollary~\ref{Cor:GES_convex} and Theorem~\ref{Thm:InvarinceGAS}. Based on $K$ constructed by the effective numerical approach proposed in Appendix~\ref{App:Compute_K}, we can obtain the value of the constant $c=0.0232$ in the assumption \textbf{A1.2}. 
{The simulation results are shown as graphs of the trace norm distance of the actual state from the target state $\frac12\|\rho-\bar{\rho}\|_1$ against time. }
The following simulations have been run with step length $\delta t = 0.002$, number of steps $N = 25000$, and with $200$ realizations each. 
By setting $\epsilon=0.3$, we have the lower bound of the dwell time $t_D=0.016$ defined in the proof of Theorem~\ref{Thm:GES_average}. By taking the dwell time $\Delta t=5\delta t<t_D$, we have that the behaviors of the trace norm distance along sample trajectories under the switching strategies $\sigma_2$ and $\sigma_3$ are bounded by the exponential reference $\epsilon c/2$ given in Corollary~\ref{Cor:GES_Dwell} and Theorem~\ref{Thm:GES_dwell}, respectively (see Fig.\ref{SIM:Fixed}). 
For simulations of applying the hysteresis switching strategies, we set up to check the system state at each step time $\delta t$. We observe that the behaviors of the trace norm distance along sample trajectories under $\sigma_1$ and $\sigma_4$ are bounded by the exponential reference $\epsilon c/2$ given in Theorem~\ref{Thm:GES_average} and Theorem~\ref{Thm:GES_state}, respectively (see Fig.\ref{SIM:Hysteresis}). For both simulations, the measurement-based trajectories converge faster than the average-based one, which is consistent with our expectation. 
%It is worth mentioning that, for the simulation with fixed dwell time~Fig.\ref{SIM:Fixed}, we set the dwell time $t_D=2\delta t$ due to the performance of the computer, the superiority of the measurement-based switching strategies cannot be shown.  

In~\cite[Section V]{liang2022GHZ}, the authors discuss the Hamiltonian feedback control of multi-qubit systems toward the target GHZ state in presence of only $z$-type measurements, i.e., the measurement operators are in the form $\bigotimes^n_{i=1}\sigma^i$ with $\sigma^i\in\{\sigma_z,\mathbf{I}\}$. In this case, the GAS is not proved in~\cite{liang2022GHZ} since only Hamiltonian control may not drive the system through some invariant sets and approach the target state. In this paper, we combine the advantages of measurement-based feedback and dissipative control to easily overcome the above obstacles.

{
\subsection{Spin-$\frac{3}{2}$ system under assumption \emph{\textbf{A2}}}\label{sec:spin1}
In this example we illustrate the potential of the novel control strategy $\sigma_5$ on a  four-level system, so that $\mathcal{H}\simeq \mathbb{C}^{4}$. The target space is $\mathcal{H}_S = {\rm span}\{|00\rangle, |01\rangle\}$ and the initial state is $\rho_0=\frac{1}{2}(|00\rangle + |11\rangle)(\langle00|+\langle11|)$ .
We consider three Lindbladiangenerators as follows
\begin{equation*}
\begin{split}
    \mathcal{L}_1(\rho) &= -i[H,\rho]+\mathcal{D}_C(\rho);\\
    \mathcal{L}_2(\rho) &= -i[-H,\rho]+\mathcal{D}_C(\rho);\\
    \mathcal{L}_3(\rho) &= \mathcal{D}_L(\rho)+\mathcal{D}_C(\rho);
\end{split}
\end{equation*}
where the measurement$C=\mathrm{diag}(2,1,-1,-2)$ monitors the basis state population and
\begin{equation*}
    H = \left[
    \begin{matrix}
    0 & 0 & -i & 0\\
    0 & 0 & 0 & 0\\
    i & 0 & 0 & 0\\
    0 & 0 & 0 & 0
    \end{matrix}
    \right], \quad 
    L = \left[
    \begin{matrix}
    1 & 0 & 1 & 0\\
    0 & 1 & 0 & 0\\
    0 & 0 & 0 & 1\\
    0 & 0 & 0 & 0
    \end{matrix}
    \right]
\end{equation*} are associated to the controlled coherent and incoherent transitions, respectively.
One can verify that the target space $\mathcal{H}_S$ is not invariant for $\mathcal{L}_1$, $\mathcal{L}_2$ and $\mathcal{L}_3$ via Theorem~\ref{Thm:InvarinceGAS}. Thus, the switching strategies $\sigma_1$, $\sigma_2$, $\sigma_3$ and $\sigma_4$ cannot be applied. Consider $K=\mathrm{diag}(0,0,k_1,k_2)$ with $k_1,k_2>0$, by straightforward calculation, the assumption \textbf{A2} is satisfied if and only if $k_2>k_1$. In the switching strategy $\sigma_5$, the constant $\bar{M}_K$ can be estimated by 
\begin{equation*}
    \bar{M}_K\leq \max_{i\in\{1,2,3\}} \|\mathrm{Tr}({\mathcal{L}^*_i}^2(K))\|,
\end{equation*}
where $\mathrm{Tr}(\rho^2)\leq 1$ and Cauchy-Schwarz inequality are used. The following simulations have been run with step length $\delta t = 0.005$, number of steps $N = 5000$, and $500$ realizations.
By setting $k_1=1$, $k_2=2$ and the dwell time as $100\delta t$, the convergence of the system toward $\mathcal{H}_S$ under $\sigma_5$, starting at $\rho_0$, is shown in Fig.\ref{SIM:Modulate} as a graph of the average $\mathbf{d}_{\mathfrak{S}}(\rho)=||\rho-\Pi\rho\Pi||_1$ as  in Definition \ref{def:distance} where $\Pi={\rm diag}(1,1,0,0)$.
\begin{figure}[!t]
\centerline{\includegraphics[width=\columnwidth]{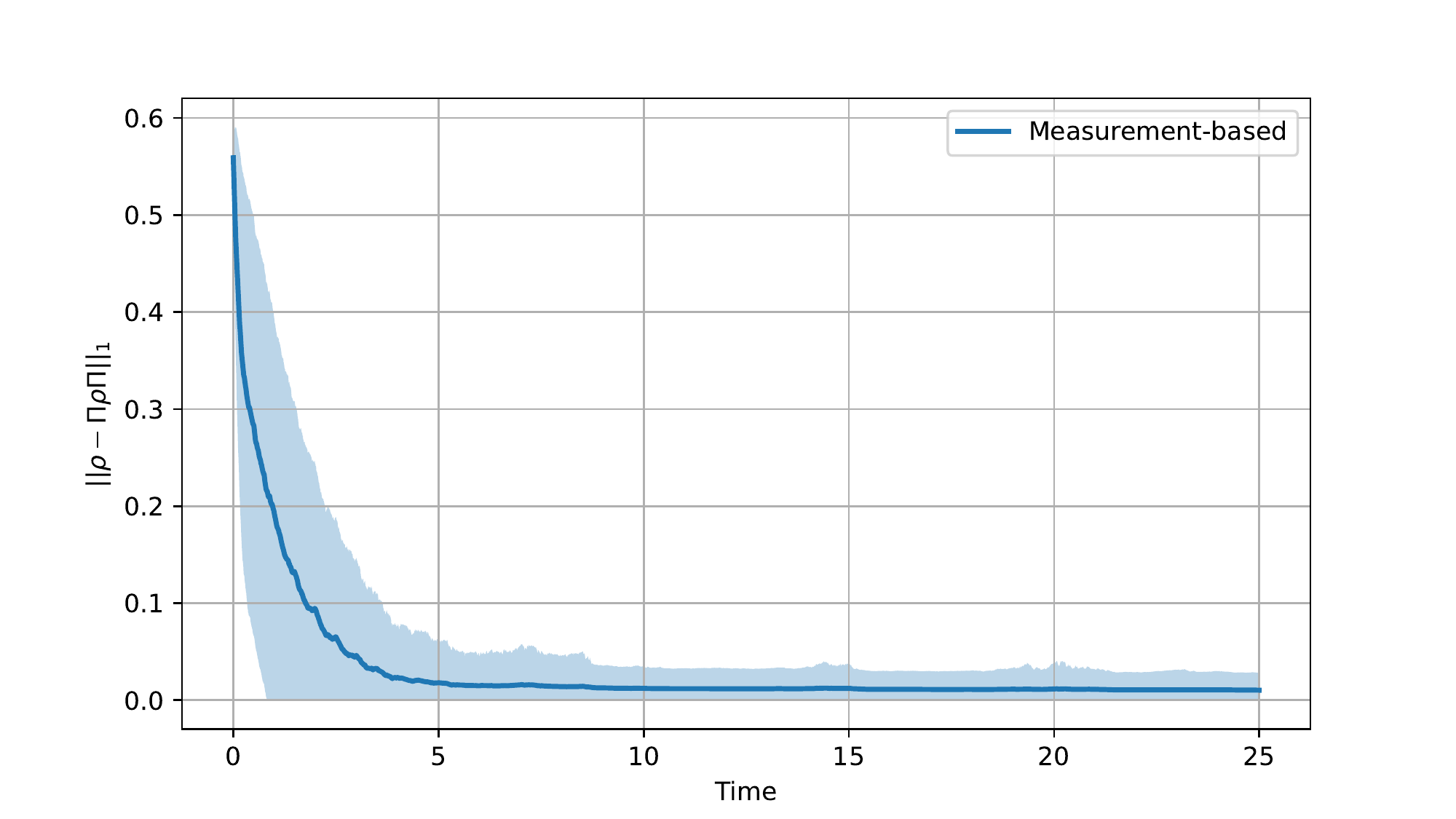}}
\caption{Asymptotic stabilization of a spin-$\frac{3}{2}$ system toward $\mathcal{H}_S$ starting from $\rho_0$ by modulating Lindbladians. The blue curve represents the mean value of 500 realizations under $\sigma_5$ and the light blue area shows the mean value plus or minus one standard deviation.}
\label{SIM:Modulate}
\end{figure} The results show how stabilization of the subspace is effectively obtained using the  proposed  control strategy.
}

\section{Conclusion}
This work presents a thorough analysis of switching techniques for fast stabilization of pure states and subspaces for quantum filtering equations. We demonstrate how globally exponentially stabilizing control laws can be derived both based on the expected  state dynamics, allowing for off-line computation and open-loop implementation, and the current state estimate conditioned on the measurement record, which leads to more accurate switching and  better performance in most cases \cite{grigoletto2021stabilization}. In proving GES, we propose ways to compute upper bounds for the convergence exponent. 

We also provide a novel way to derive feedback laws that include dissipative control actions and ensure global asymptotic stability even in the case where the controlled dynamics do not necessarily maintain the target invariant, allowing us to address stabilization problems that were so far only considered with Hamiltonian control actions, see e.g. Section \ref{sec:spin1}.
This is possible by suitably modulating the amplitude of the switching controlled dynamics, avoiding chattering and stability problems that so far allowed to prove only weaker results. Because of the modulation,  GAS can be proved  but in general we lose exponential convergence. These results thus open new possibilities in the integration of coherent and dissipative resources for the control of quantum systems, and extend the applicability of feedback methods for effectively controlling quantum systems. 

Further developments of this line of work are possible. We aim to  derive better bounds for the exponential convergence speed, as the simulation results indicate that the proposed ones are not always tight. This is likely due the fact that the  stochastic back-action component of the dynamics is not directly taken into account in computing such bounds. Other potential developments include the robustness of the filters and the proposed control law to initialization errors, simplified switching laws that do not need a full state reconstruction, and potential experimental applications.

\appendix
\section{Instrumental Results}

Here, we state an instrumental result that is used in the proof of the GES of the target subspace almost surely.
\begin{lemma}
Suppose that \emph{\textbf{A1.1}} holds true.
For all initial state $\rho_0\in\mathcal{S}(\mathcal{H})\setminus\mathcal{I}(\mathcal{H}_S)$, 
$
\mathbb{P}\big( \rho(t)\in \mathcal{S}(\mathcal{H})\setminus\mathcal{I}(\mathcal{H}_S),\,\forall t\geq0\big)=1,
$
where $\rho(t)$ is a solution of the switched system~\eqref{Eq:SSME}.
\label{Lemma:NeverReach}
\end{lemma}
\emph{Proof.}
Let us consider the function $\mathrm{Tr}(\Pi_{\mathfrak{R}}\rho)=\mathrm{Tr}(\rho_R)\in[0,1]$, $\mathrm{Tr}(\Pi_{\mathfrak{R}}\rho)=0$ if and only if $\rho\in\mathcal{I}(\mathcal{H}_S)$. 
Then, it is sufficient to show that $\mathbb{P}\big( \mathrm{Tr}(\Pi_{\mathfrak{R}}\rho_(t))>0,\,\forall t\geq0\big)=1$ for all $\rho_0\in\mathcal{S}(\mathcal{H})\setminus\mathcal{I}(\mathcal{H}_S)$.
We follow the similar arguments as in~\cite[Proposition 4.4]{benoist2017exponential}. Let
\begin{equation*}
\rho_{R,red}=
\begin{cases}
\frac{\rho_R}{\mathrm{Tr}(\rho_R)},& \text{if }\mathrm{Tr}(\rho_R)\neq 0;\\
\mu_R,& \text{if }\mathrm{Tr}(\rho_R)= 0,
\end{cases}
\end{equation*}
where $\mu_R\in\mathcal{S}(\mathcal{H}_R)$ is arbitrary. Due to It\^o formula~\cite[Theorem 2.32]{protter2004stochastic}, we have the following Doleans-Dade exponential
\begin{align*}
&d\mathrm{Tr}(\Pi_{\mathfrak{R}}\rho(t))\\
&=\mathrm{Tr}(\Pi_{\mathfrak{R}}\rho(t))\Big(\sum^{m}_{k=1}u^k_t\mathrm{Tr}\big(\mathcal{L}_{k,R}\rho_{R,red}(t)\big)dt+\\
&~~~\big(\mathrm{Tr}((C_R+C_R^*)\rho_{R,red}(t))- \mathrm{Tr}((C+C^*)\rho(t) \big)dW(t)\Big).
\end{align*}
It can be also written in the following form, for any $t\geq 0$,
{\small
\begin{align*}
&\mathrm{Tr}(\Pi_{\mathfrak{R}}\rho(t))\\
&=\mathrm{Tr}(\Pi_{\mathfrak{R}}\rho_0)\exp\Big\{\int^{t}_{0}\sum^{m}_{k=1}u^k_s\mathrm{Tr}\big(\mathcal{L}_{k,R}\rho_{R,red}(s)\big)ds\\
&-\frac{1}{2}\int^{t}_{0}\big(\mathrm{Tr}((C_R+C_R^*)\rho_{R,red}(s))- \mathrm{Tr}((C+C^*)\rho(s) \big)^2ds\\
&+\int^{t}_{0}\big(\mathrm{Tr}((C_R+C_R^*)\rho_{R,red}(s))- \mathrm{Tr}((C+C^*)\rho(s) \big)dW(s)\Big\}.
\end{align*}
}
Thus, the strict positivity of $\mathrm{Tr}(\Pi_{\mathfrak{R}}\rho(t))$ is clear and then the proof is conclude.
\hfill$\square$

\medskip

The following lemmas are used to investigate the switching strategy in Section~\ref{Sec:Modulate}.
\begin{lemma}
For any $K\in\mathcal{B}(\mathcal{H})$, 
$$\mathbf{L}_K(\rho):=\min_{k\in\mathcal{M}}\Tr(K\mathcal{L}_k(\rho))$$ is continuous in $\mathcal{S}(\mathcal{H})$.
\label{Lem:Continuity}
\end{lemma} 
\emph{Proof.}
Denote $j_a:=\arg\min_{k\in\mathcal{M}}\Tr(K\mathcal{L}_k(\rho_a))$ and $j_b:=\arg\min_{k\in\mathcal{M}}\Tr(K\mathcal{L}_k(\rho_b))$. One deduces
\begin{align*}
|\mathbf{L}_K(\rho_a)-\mathbf{L}_K(\rho_b)|\leq &\max\big\{\big|\Tr\big(K (\mathcal{L}_{j_a}(\rho_a)-\mathcal{L}_{j_a}(\rho_b)) \big)\big|,\\
&~~~~~~~~\big|\Tr\big(K (\mathcal{L}_{j_b}(\rho_a)-\mathcal{L}_{j_b}(\rho_b)) \big)\big|\big\}.
\end{align*}
Moreover, for all $j\in\mathcal{M}$, $\mathcal{L}_j(\rho)$ is continuously differentiable in the each matrix element of $\rho$. Since the compactness of $\mathcal{S}(\mathcal{H})$, the derivatives of $\mathcal{L}_j(\rho)$ with respect to the matrix elements of $\rho$ are bounded. Thus, $\mathcal{L}_j(\rho)$ is Lipschitz continuous, i.e., there exists a constant $c>0$ such that $\|\mathcal{L}_j(\rho_a)-\mathcal{L}_j(\rho_b)\|\leq c\|\rho_a-\rho_b\|$ for any $\rho_a,\rho_b\in\mathcal{S}(\mathcal{H})$. Due to the Cauchy-Schwarz inequality, we have 
\begin{align*}
\big|\Tr\big(K (\mathcal{L}_j(\rho_a)-\mathcal{L}_j(\rho_b)) \big)\big|&\leq \|K\|\|\mathcal{L}_j(\rho_a)-\mathcal{L}_j(\rho_b)\| \\
&\leq d\|\rho_a-\rho_b\|,
\end{align*}
for some constant $d>0$. The proof is complete.
\hfill$\square$

\begin{lemma}
Suppose that $\mathbf{L}_K(\rho)<0$ for all $\rho\in\mathcal{S}(\mathcal{H})\setminus\mathcal{I}(\mathcal{H}_S)$ and $K$ is the extension in $\mathcal{B}(\mathcal{H})$ of $K_R\in\mathcal{B}_{>0}(\mathcal{H}_R)$. Then, $\mathbf{L}_K(\rho)=0$ for all $\rho\in\mathcal{I}(\mathcal{H}_S)$.
\label{Lem:Min=0}
\end{lemma}
\emph{Proof.}
We prove this lemma by contradiction for the case $\mathrm{dim}(\mathcal{H}_S)\geq 2$, the case $\mathrm{dim}(\mathcal{H}_S)=1$ can be done in the same manner. Due to Lemma~\ref{Lem:Continuity}, $\mathbf{L}_K(\rho)$ is continuous on $\mathcal{S}(\rho)$. We suppose that there exists $j^*\in\mathcal{M}$, a constant $\delta>0$ and a non-empty open subset $D\subset \mathcal{I}(\mathcal{H}_S)$ such that $\Tr(K\mathcal{L}_{j^*}(\rho))<-\delta$ for all $\rho\in D$. Note that $\Tr(K\rho)\geq0$ for all $\rho\in\mathcal{S}(\mathcal{H})$ and $\Tr(K\rho)=0$ if and only if $\rho\in\mathcal{I}(\mathcal{H}_S)$. Consider the following Lindbladian equation
$$
\frac{d}{dt}\rho(t)=\mathcal{L}_{j^*}(\rho(t)), \quad \rho(0)\in D.
$$
For a sufficiently small $\epsilon>0$, $\rho(t)\in D$ due to the continuity. Then, we have
\begin{equation*}
\Tr(K\rho(\epsilon))=\Tr(K\rho(0))+\int^{\epsilon}_0\Tr(K\mathcal{L}_{j^*}(\rho(s)))ds<-\delta \epsilon,
\end{equation*}
which contradicts to the fact $\Tr(K\rho)\geq 0$ for all $\rho\in\mathcal{S}(\mathcal{H})$.
\hfill$\square$

\section{Numerical scheme for constructing $K$}
\label{App:Compute_K}
Suppose there exists $\gamma\in\{\gamma\in(0,1)^m| \sum_{j\in\mathcal{M}}\gamma_j=1\}$ such that $\mathcal{H}_S$ is GAS for the Lindbladian 
$
\mathcal{L}_{\gamma}(\rho)=\sum_{j\in\mathcal{M}}\gamma_j\mathcal{L}_{j}(\rho)
$. Then, the spectral abscissa $\alpha_\gamma$ of $\mathcal{L}_{\gamma,R}$ is strictly positive. If $\mathcal{L}^*_{\gamma,R}$ generates a semigroup of irreducible maps, then there exists $K_R\in\mathcal{B}_{>0}(\mathcal{H})$ such that $\mathcal{L}^*_{\gamma,R}(K_R)=-\alpha_\gamma K_R$ due to Perron-Frobenius theorem~\cite{evans1977spectral}. Otherwise, there exists $K_R\in\mathcal{B}_{*}(\mathcal{H})$ such that $\mathcal{L}^*_{\gamma,R}(K_R)=-\alpha_\gamma K_R$ since $\mathcal{L}^*_{\gamma,R}$ is a Hermitian preserving linear map.
In the following, we propose a numerical approach to construct $X\in\mathcal{B}_{>0}(\mathcal{H})$ such that $\mathcal{L}^*_{\gamma,R}(X)\leq -c X$ with $c>0$ based on the eigenvector $K_R\in\mathcal{B}_{*}(\mathcal{H})$ of $\mathcal{L}^*_{\gamma,R}$ corresponding to the eigenvalue $-\alpha_\gamma$. The existence of such $X$ and $c$ is proved by Theorem~\ref{Thm:SpectralAbscissa}. 

In order to better employ the existing algorithm of computing eigenvectors and eigenvalues, we associate a $N^2$-dimensional vector $v$ to a $N\times N$-dimensional matrices by a Hilbert-Schmidt orthonormal basis of the set of Hermitian matrices, e.g., the identity matrix $\Phi_0 = \mathbf{I}$ and the set of $N^2-1$ extended Gell-Mann matrices ${\Phi_j}$ where $\Phi_j=\Phi_j^*$, $\mathrm{Tr}(\Phi_j)=0$ and $\mathrm{Tr}(\Phi_j\Phi_i)=\delta_{i,j}$. Thus, the linear map $\mathcal{L}^*_{R}$ can be represented as a matrix $M$ on the basis $\{\Phi_j\}^{N^2-1}_{j=0}$. Then, all matrix representations on the basis $\{\Phi_j\}$ of eigenvectors $X_i\in\mathcal{B}_{*}(\mathcal{H}_R)$ with $i\in\{1,\dots,n\}$ associated to the eigenvalue $-\alpha_\gamma$ can be computed explicitly, where $n$ denotes the geometric multiplicity of $-\alpha_\gamma$.
From a geometrical point of view, all positive definite matrices define a convex cone, we can determine a $X(\bar{\beta}):=\sum_i\bar{\beta}_i X_i$ for $\bar{\beta}\in\mathbb{R}^n$ with unit norm such that the angle between $X(\bar{\beta})$ and a fixed positive definite matrix $A>0$ is the largest in Hilbert-Schmidt inner product, one then solves an optimization problem:
\begin{equation*}
   \bar{\beta} := \operatorname*{arg\,max}_{\beta \in \mathbb{R}^n} \frac{\mathrm{Tr}(A X(\beta))}{\sqrt{\mathrm{Tr}(X(\beta)^2)}}.
\end{equation*}
For the sake of simplicity, we set $A=\mathbf{I}$. Since extended Gell-Mann matrices are trace-less, we have
\begin{equation*}
   \bar{\beta} = \operatorname*{arg\,max}_{\beta \in \mathbb{R}^n} \frac{\sum^n_{k=1} \beta_k\mathrm{Tr}(X_k)}{\sqrt{\sum^n_{k=1} \beta_k^2}}=\operatorname*{arg\,max}_{\|\beta\|=1}\sum^n_{k=1} \lambda_k\mathrm{Tr}(X_k),
\end{equation*}
which implies
\begin{equation*}
    \bar{\beta}_k = \frac{\mathrm{Tr}(X_k)}{\sqrt{\sum^n_{i=1}\mathrm{Tr}(X_i)^2}},\quad k=\{1,\dots,n\}.
\end{equation*}
Then, the constructed $X(\bar{\beta})\in\mathcal{B}_{*}(\mathcal{H}_R)$ is either positive definite or close to the positive definite cone. If it is not positive definite, we need to perturb the non-positive spectral of $X(\bar{\beta})$ so that the perturbed $\tilde{X}(\bar{\beta})>0$ and $\mathcal{L}^*_{\gamma,R}(\tilde{X}(\bar{\beta}))<0$. Then, we have $c=\bar{\lambda}(\mathcal{L}^*_{\gamma,R}(\tilde{X}(\bar{\beta})))/\underline{\lambda}(\tilde{X}(\bar{\beta}))$, where $\bar{\lambda},\underline{\lambda}$ denote the maximum and minimum eigenvalues respectively.

\bibliographystyle{plain}
\bibliography{Ref_Switching}

\end{document}